\documentclass[12pt]{scrartcl}
\usepackage{listings, xcolor}
\usepackage{amsmath}
\usepackage{amssymb}
\usepackage{graphicx}
\usepackage{caption}
\usepackage{subcaption}
\usepackage[T1]{fontenc}
\usepackage[latin1]{inputenc}
\usepackage[english]{babel}
\usepackage[latin1]{inputenc}
\usepackage{lmodern}
\usepackage{bbm}
\usepackage{float} 
\usepackage{epstopdf}
\usepackage{subcaption}

\usepackage{threeparttable}
\usepackage{setspace}
\usepackage[round]{natbib}
\usepackage{babelbib}
\usepackage{url}
\usepackage{rotating}
\usepackage{dcolumn}

\usepackage{booktabs}
\usepackage{tabularx}
\usepackage{indentfirst}
\usepackage{multirow}
\usepackage{framed}

\usepackage{chngcntr}
\counterwithin{figure}{section}
\counterwithin{table}{section}
\counterwithin{equation}{section}
\usepackage{setspace}

\usepackage{colortbl}

\usepackage[left=3cm,right=3cm,top=2cm,bottom=2cm,includeheadfoot]{geometry}
\usepackage{scrlayer-scrpage}
\usepackage{ragged2e}
\usepackage[overload]{empheq}
\ofoot[\pagemark]{\pagemark}
\color{black}
\usepackage[titletoc,title]{appendix}

\usepackage[hidelinks,breaklinks=true]{hyperref}

\newcommand{\indep}{\rotatebox[origin=c]{90}{$\models$}}

\newtheorem{theorem}{Theorem}[section]

\newtheorem{prop}{Proposition}[section]

\usepackage{dcolumn}
\newcolumntype{d}[1]{D{.}{.}{#1} }
\newcolumntype{d}[1]{D{.}{\cdot}{#1} }
\newcolumntype{Y}{>{\centering\arraybackslash}X}
\newcolumntype{Z}{>{\flushleft\arraybackslash}X}

\newcommand{\titleinfo}{\vspace{-3cm} Efficient Covariate Balancing for the Local Average Treatment Effect}
\title{\titleinfo}
\def\authora{Phillip Heiler}
\def\aff{Aarhus University} 
\def\emaila{\href{mailto:pheiler@econ.au.dk}{pheiler@econ.au.dk}}

\begin{document}
	\begin{titlepage}
		\title{\titleinfo \thanks{ \scriptsize I would like to thank Michael Knaus, Winfried Pohlmeier, Julian Sch\"ussler, Qingyuan Zhao, Toru Kitagawa, and the participants of the Rotterdam Winter Meeting of the Econometric Society 2019 and the St.~Gallen Causal Machine Learning Workshop 2020 for fruitful discussions and comments that helped to greatly improve the paper. The author gratefully acknowledges the financial support from the German Research Foundation through Project 219805061 and by the Graduate School of Decision Sciences (University of Konstanz). All remaining errors are mine.}}
		\author{\authora\thanks{{\scriptsize Department of Economics and Business Economics, CREATES, TrygFonden's Centre for Child Research, Fuglesangs Allé 4, 8210 Aarhus V, Danemark, email: \emaila.}} \\ \aff }

		\date{{{\today}} \\}
		\maketitle
		\thispagestyle{empty}
		
		\begin{abstract} \singlespacing	\small
			This paper develops an empirical balancing approach for the estimation of treatment effects under two-sided noncompliance using a binary conditionally independent instrumental variable. The method weighs both treatment and outcome information with inverse probabilities to produce exact finite sample balance across instrument level groups. It is free of functional form assumptions on the outcome or the treatment selection step. By tailoring the loss function for the instrument propensity scores, the resulting treatment effect estimates exhibit both low bias and a reduced variance in finite samples compared to conventional inverse probability weighting methods. The estimator is automatically weight normalized and has similar bias properties compared to conventional two-stage least squares estimation under constant causal effects for the compliers. We provide conditions for asymptotic normality and semiparametric efficiency and demonstrate how to utilize additional information about the treatment selection step for bias reduction in finite samples. The method can be easily combined with regularization or other statistical learning approaches to deal with a high-dimensional number of observed confounding variables. Monte Carlo simulations suggest that the theoretical advantages translate well to finite samples. The method is illustrated in an empirical example. 
		\end{abstract}
	\end{titlepage}

\newpage \pagenumbering{arabic}
\section{Introduction}
Estimation of causal effects is at the heart of modern empirical research in economics. In particular, evaluating the impact of policies or programs on units with heterogeneous preferences and characteristics such as households, workers, unemployed, firms, or students is necessary to develop a thorough understanding of fundamental economic relationships. This paper deals with the problem of estimating a causal effect of a treatment using variation from a conditionally independent instrumental variable when units have heterogeneous responses to the instrument and to the treatment. 
We develop a semiparametric estimation method for the local average treatment effect (LATE) based on inverse probability weighting (IPW).
The method is designed to increase internal validity by reducing the finite sample bias in estimation of the treatment effect while allowing for dependent observable and unobservable variables to affect both treatment participation and potential outcomes of interest (\textit{selection on unobservables}).
It is more robust than conventional instrumental variable methods as it does not require parametric functional form assumptions about outcome or treatment selection steps or other restrictions such as constant causal effects. Moreover, it has appealing point estimation properties compared to conventional IPW estimators for the LATE.


The key insight required is that IPW estimation of the LATE fundamentally rests on \textit{balancing} the distribution of observable confounders for two reduced form type components. In particular, for the reduced form type component of the outcome, observations that are instrumented at a given level are weighted by their inverse probability of being in that state, i.e.~by their \textit{instrument propensity scores}. This assures that differences in observed confounders are correctly taken into account when estimating the reduced form type effect for units that choose treatment in accordance with the instrument (compliers). For the overall LATE, the remaining differences due to heterogeneous treatment selection responses to the instrument are then taken into account by another reduced form type component that applies the same inverse probability weights to the corresponding treatment indicators.

On a technical level, balancing means that the empirical distributions of inverse probability weighted observable covariates between the groups that are defined by the instrument are imposed to be identical. However, conventional methods for estimation of the weights such as maximum likelihood or even true weights do not yield perfect balance in finite samples. We exploit recent advances in the literature on the evaluation of causal effects that have been established under the more restrictive conditionally independent treatment and overlap assumptions (\textit{selection on observables}) to construct empirical balancing conditions for the estimation of inverse probability weights. 
We achieve exact finite sample balance through tailoring the loss function for the instrument propensity scores that are used for estimation of the LATE. 
The method compares favorable to conventional inverse probability weighting methods as the tailored loss approach minimizes approximate bias while simultaneously favoring weights that do not exhibit too much variance. In addition, it preserves the design philosophy by \cite{rubin2007design} as it does not require the use of any outcome or treatment data when selecting a model for the instrument propensities which helps to avoid p-hacking and other post-model-selection problems.  

Throughout the paper we show that the proposed balancing estimator has desirable bias properties that compare to the conventional two-stage least squares estimator under homogeneous causal effects for the compliers. Moreover, the bias can be further reduced by incorporating additional information into the balancing constraints, in particular from the treatment selection step. 
The balancing approach implicitly penalizes deviations from moderate inverse probability weights and thus helps to reduce the variance in estimation of the LATE. Moreover, the balancing estimator is asymptotically normal and reaches the semiparametric efficiency bound if the number of balancing constraints grows appropriately with the sample size. The method can be easily combined with regularization or other statistical learning approaches to deal with a high-dimensional number of observed confounding variables. Monte Carlo simulations suggest that the theoretical advantages over conventional methods translate well to finite samples. The method is applied to a re-evaluation of the causal effect of 401(k) participation on total financial assets.    

Identification and non/semiparametric estimation of the LATE under a binary conditionally independent instrument has been first considered by \cite{abadie2003semiparametric} and \cite{frolich2007nonparametric}. \cite{abadie2003semiparametric} relies on implicit identification of complying units by inverse probability weights to estimate or approximate conditional expectation functions for compliers that can consequently be used to construct estimates for local average treatment effects. His method requires an (approximate) model for complier outcomes and is thus more prone to misspecification and post-model-selection problems. In practice, the method often yields estimates close or identical to conventional two-stage least squares if linear models are used \citep{angrist2009mostly}.  \cite{frolich2007nonparametric} proposes nonparametric imputation/matching estimators for estimation of the LATE and shows that the semiparametric efficiency bound is not affected by knowledge of the instrument propensity scores similar to \cite{hahn1998role} in the context of selection on observables. \cite{frolich2007nonparametric} also suggests the use of an IPW estimator but does not provide any theory for estimation. \cite{donald2014testing} close this gap in the literature by using an IPW estimator for the LATE that relies on nonparametric series estimation for the instrument propensity scores similar to \cite{hirano2003efficient} in the context of selection on observables and provide conditions for semiparametric efficiency. \cite{donald2014inverse} suggest to estimate the LATE semiparametrically efficient via IPW with instrument propensity scores estimated by local polynomial regression and provide a higher order mean squared error expansion. The IPW estimators in \cite{frolich2007nonparametric}, \cite{donald2014inverse}, and \cite{donald2014testing} are all of the ``IPWI''-type, i.e.~do not impose normalization of the inverse probability weights. Our approach is closest to \cite{donald2014testing} but uses balancing conditions instead of a series logistic estimator. Moreover, our balancing estimator is automatically weight-normalized, favors moderate instrument propensities, and imposes exact mean balance in finite samples. Thus, despite their asymptotic equivalence, we expect major differences in bias and overall point estimation risk in finite samples. 

Improving IPW estimation through imposing exact or approximate balancing constraints has been considered in the literature on estimation of treatment effect under the more restrictive selection on observables assumptions. \cite{graham2012inverse} considers tilted moment conditions for estimation of the conventional propensity scores that bear close resemblance to balancing approaches. \cite{hainmueller2012entropy} proposes direct optimization of a distance criterion depending on the inverse probability weights (e.g. Kullback entropy divergence) subject to approximate empirical balancing and positivity constraints. \cite{zhao2017entropy} provide conditions under which the balancing method by \cite{hainmueller2012entropy} is doubly robust for the average treatment effect on the treated. \cite{imai2014covariate} consider exact balancing of covariates through propensity scores in a parametric GMM or empirical likelihood framework. \cite{zubizarreta2015stable} proposes to minimize a quadratic problem in terms the inverse probability weights subject to similar approximate mean balancing constraints. \cite{athey2018approximate} employ approximate balancing weights for bias-correction of a high-dimensional linear model for estimation of treatment effects. \cite{zhao2019covariate} develops a unifying framework for empirical balancing approaches and demonstrates how to tailor the (negative) loss function to produce weights that correspond to the treatment effect of interest. Moreover, \cite{zhao2019covariate} proposes methods for regularization and shows how the bias is affected by relaxing exact balancing conditions to approximate balancing. All of these contributions suggest that, for selection on observables, empirical balancing can substantially outperform conventional weighting estimators by reducing differences between the weighted empirical distributions of treatment and control units. Our balancing approach in its most basic form is a repeated application of the exact balancing method by \cite{imai2014covariate} and \cite{zhao2019covariate} using the instrument instead of the treatment indicators. However, it has different bias and weight normalization properties that do not apply in the context of selection on observables. Moreover, for the LATE there is an extended hierarchy in terms of the available information that consists of three different levels: The basic instrument assignment and the higher-order treatment selection and outcome generation steps. We demonstrate that using information from the higher-order steps can help to achieve approximately unbiased estimates for the causal effect. In particular, balancing estimated treatment participation probabilities for a given instrument level can help to reduce point estimation risk even when the treatment selection model is misspecified. 

Applications that rely on identification of a conditionally independent instrument are manifold. \cite{angrist1990lifetime} studies the effect of military service on lifetime earnings using the Vietnam era draft lottery as an instrument. \cite{card1995using} uses college proximity as an instrument for education, see also \cite{card2001estimating}. \cite{poterba1995401} and \cite{abadie2003semiparametric} exploit eligibility to 401(k) programs as an instrument to investigate whether contributions to these plans crowd out other personal savings. \cite{cawley2007impact} and \cite{cawley2013impact} use minimum requirements for physical education (PE) as an instrument for actual time spend in PE to evaluate the impact on health outcomes for elementary school children and high school students respectively. Based on individual data from different federal states in Germany, \cite{knaus2018better} exploit between and within variation of compulsory PE lessons as an instrument to estimate the effects of PE on multiple measurements for child development. The working paper by \cite{knaus2018better} is an application of an idea similar to the one proposed in this paper. They combine two steps of the inverse probability tilting method by \cite{graham2012inverse} to obtain a balanced estimate for the LATE. However, they do not provide any theory for estimation or statistical inference. 

The paper is structured as follows: Section \ref{sec_BAL_identification} introduces the identification assumptions and provides the basic arguments behind the balancing approach. Section \ref{sec_BalEstimation} presents the estimation strategy. Section \ref{sec_BAL_properties} provides the statistical properties of the balancing estimator. Section \ref{sec_BAL_extensions} demonstrates how to use higher-order information for balancing and outlines extensions to the case of high-dimensional observed confounding variables. Section \ref{sec_BalMC1} contains the Monte Carlo simulations. Section \ref{sec_ApplicationLATE} provides the application and Section \ref{sec_BAL_conclusion1} concludes. All major proofs and derivations are collected in the Appendix.  

\section{Identification and Balancing} \label{sec_BAL_identification}
We consider standard identification conditions for treatment effects under unobservable heterogeneity and a binary conditionally independent instrument often referred to as \textit{assignment}. Assume we observe independent data $(Y_i,D_i,Z_i,X_i')$ for units $i=1,\dots,n$. $X_i$ is a vector of (causally) predetermined\footnote{Predetermined variables are set before the cause can have an effect, e.g.~a priori attributes, see \cite{imbens2015causal}.} random variables supported on $\mathcal{X} \subset \mathbb{R}^{\dim{X_i}}$, $Z_i$ a binary instrument, $D_i$ a binary treatment indicator, and $Y_i$ a real-valued outcome variable. In principle, there are four potential outcome states $Y_i(d,z)$ for $d,z \in \{0,1\}$ but only $Y_i = Y_i(D_i,Z_i)$ is observed. For each instrument level $z\in \{0,1\}$, there is a potential treatment status $D_i(z)$ which yields the observed treatment $D_i = D_i(Z_i)$ according to \begin{align}
D_i = Z_iD_i(1) + (1-Z_i)D_i(0).
\end{align}   
The following identification assumptions along the lines of \cite{abadie2003semiparametric} and \cite{frolich2007nonparametric} are imposed, see also \cite{donald2014inverse,donald2014testing}: 

\textit{\begin{enumerate}
		\item[A.1)] (Conditional Independence) $\{Y_i(d,z);\forall d,z\}, D_i(1), D_i(0) \ \indep \ Z_i|X_i$.
		\item[A.2)] (Exclusion) $P(Y_i(d,1) = Y_i(d,0)) = 1 \text{ for } d=0,1$.
		\item[A.3)] (Monotonicity) $ P(D_i(1) - D_i(0) \geq 0) = 1$.
		\item[A.4)] (First Stage) $E[D_i(1) - D_i(0)] \neq 0$.
		\item[A.5)] (Strong Instrument Overlap) Let $\pi(x) = P(Z_i=1|X_i=x)$. There exists a $\delta > 0$ such that $\delta < \pi(x) < 1-\delta$ for all $x \in \mathcal{X}$.
		\item[A.6)] (Stable Unit Treatment Values) $(\{Y_i(d,z);\forall d,z\}, D_i(1), D_i(0))$, $i=1,\dots,n$ are independent for $i\neq j$. 
	\end{enumerate}}
	A.1-A.4 together are a weaker version of the conditions for identification of the LATE in the seminal paper by \cite{imbens1994identification} that rely on a completely independent instrument. Assumption A.1 is the fundamental identification assumption. It implies that after controlling for a sufficient set of observed confounding variables, all residual variation in potential outcomes and potential treatment statuses is completely independent of the instrument. Thus, conditional on observed confounders, the instrument can be thought of as being allocated like in a completely randomized experiment. Assumption A.2 rules out any direct effects of the instrument on potential outcomes other than through the indirect treatment channel. Thus, there cannot be any unobserved confounders affected by the instrument and no feedback from potential outcomes to the instrument. Under this exclusion, the fundamental problem of causal inference \citep{holland1986statistics} determines the observation rule for the outcome, i.e. \begin{align} Y_i &= D_iY_i(1,Z_i) + (1-D_i)Y_i(0,Z_i) \notag \\ &= D_iY_i(1) + (1-D_i)Y_i(0).  \end{align} Assumption A.3 imposes a monotonous effect of the instrument on the treatment choice, i.e.~receiving instrument $Z_i = 1$ makes any unit at least as likely to select itself or to be selected into treatment compared to $Z_i = 0$. This is sometimes referred to as the ``no defiers'' assumption, see \cite{angrist1996identification}. Assumption A.4 assures that there is overall variation in potential treatment statuses as a result from variation in the instrument. Together with monotonicity, this requires the data to have a nonzero share of compliers, i.e.~there must be units for which $D_i(1) > D_i(0)$. Thus, the population cannot only be comprised of units that are always treated or never treated independently of their instrument level. Assumption A.5 requires that potentially each unit could have been exposed to a different instrument level. In principle, point identification only requires instrument overlap ($\delta = 0$). However, for regular behavior of the point estimators considered throughout the paper, strong instrument overlap is required.\footnote{Irregular estimators for the case of identification under selection on observables are considered in  \cite{heiler2020inference}.} 
	Assumption A.6 rules out any spillover or general equilibrium effects in terms of both potential treatment states and potential outcomes. This, together with the assumption that covariates are predetermined, allows for a coherent definition of individual causal effects as differences in potential outcomes \citep{rubin1974estimating,imbens2015causal}. The causal effect of the treatment on the outcome for a unit $i$ is then given by \begin{align}
	\tau_i = Y_i(1) - Y_i(0).
	\end{align}Note that there are no further restrictions on the functional relationship between a unit's potential outcomes, instrument, or covariates. Thus, the framework allows for almost any type of observable and unobservable heterogeneity in potential outcomes and causal effects, e.g.~it allows for nonconstant treatment effects even conditional on the observed confounders. 
	\cite{frolich2007nonparametric} shows that under similar assumptions as A.1-A.6, the local average treatment effect (LATE) \begin{equation}
	\tau_{LATE} = E[Y_i(1)-Y_i(0)|D_i(1) > D_i(0)]
	\end{equation} is nonparametrically identified. $\tau_{LATE}$ is the average treatment effect for the subpopulation of compliers, i.e.~the expected causal effect for a unit randomly drawn from the population of units that alter their potential treatment choice in accordance with the instrument. Without further assumptions, identification does not extend to more general causal effects such as the average treatment effect (ATE) or the treatment effect on the treated (TT). Under treatment effect homogeneity, the LATE is equal to the ATE and the TT. Moreover, in the case of one-sided noncompliance, i.e.~$P(D_i(0)=1) = 0$, the LATE equals the TT \citep{bloom1984accounting,frolich2013identification}. One-sided noncompliance often occurs in randomized field experiments with imperfect compliance if the treatment can only be provided by the experimenter. 
	
	The identification results in \cite{abadie2003semiparametric} and \cite{frolich2007nonparametric} allow for the construction of matching, model-based imputation, and inverse probability weighting estimators for the LATE. In this paper we focus on the latter as they only require estimation of the instrument propensity scores $\pi(X_i)$ and no information from outcome or treatment selection steps. Both treatment selection and in particular potential outcome mechanisms can be more difficult to model as they are often the results of complicated mechanisms such as markets, search and matching processes, or social and biological structures and interactions. Thus, focusing on the instrument or \textit{assignment phase} avoids misspecification and post-model-selection problems with statistical inference as outcome and treatment data are not used for obtaining the instrument propensities. This is along the lines of the argument for focusing on the \textit{design phase} of experimental and observational studies made by \cite{rubin2007design}. 
	
	In the following we outline the balancing principle behind inverse probability weighting and show how this allows us to extract population constraints that can be used for estimation of the instrument propensity scores. For exploiting identification via inverse probability weighting note that \begin{align}
	E[Y_i(1) - Y_i(0)|D_i(1) > D_i(0)] = \frac{\Delta}{\Gamma} \label{eq_LATEisRATIO1}
	\end{align}
	with \begin{align}
	\Delta &= E\bigg[\frac{Z_iY_i}{\pi(X_i)}\bigg] - E\bigg[\frac{(1-Z_i)Y_i}{1-\pi(X_i)}\bigg] \\
	\Gamma &= E\bigg[\frac{Z_iD_i}{\pi(X_i)}\bigg] - E\bigg[\frac{(1-Z_i)D_i}{1-\pi(X_i)}\bigg]. 
	\end{align}
	Thus both numerator and denominator of the LATE can be written as a difference of two expectations of inverse probability weighted quantities that can be identified from the joint distribution of $(Y_i,D_i,Z_i,X_i')$. The fundamental mechanism behind identification via inverse probability weighting is the \textit{balancing property}, i.e.~the inverse probability weights will balance the distribution of any function of covariates across instrument levels.
	Let $f_x(x)$ denote the density of the observed confounders and $f_{x|z=1}(x)$ the density of the observed confounders conditional on $Z_i=1$. By Bayes' Law it follows that for all $x \in \mathcal{X}$ \begin{align}
	\frac{f_{x|z=1}(x)}{\pi(x)}P(Z_i=1) = f_x(x).  \label{eq_BalancingDensity1}
	\end{align} 
	Thus, under Assumptions A.1 and A.5 we have that  \begin{align}
	E\bigg[\frac{Z_iD_i}{\pi(X_i)}\bigg] &= E\bigg[\frac{D_i(1)}{\pi(X_i)}\bigg|Z_i=1\bigg]P(Z_i=1) \notag \\
	&= \int_{\mathcal{X}}\frac{E[D_i(1)|X_i=x]}{\pi(x)}f_{x|z=1}(x)dx P(Z_i=1) \notag \\
	&= \int_{\mathcal{X}}E[D_i(1)|X_i=x]f_x(x)dx \notag \\
	&= E[D_i(1)]. \label{eq_BAL_ipwidentif1}
	\end{align}
	Note that the conditional mean of the observed treatment status for the units with $Z_i=1$ receives a weight such that it corresponds to the conditional mean of the potential treatment status (normalized) over the full population. Equivalently, it holds that ${(1-\pi(x))}^{-1}{f_{x|z=0}(x)}P(Z_i=0) = f_x(x)$. Thus, similar derivations as in \eqref{eq_BAL_ipwidentif1} can be done for all components in \eqref{eq_LATEisRATIO1}. In general,  property \eqref{eq_BalancingDensity1} of the inverse probability weights implies a \textit{mean balancing} for any function of the covariates: Let $g:\mathcal{X}\rightarrow\mathbb{R}$ be a measurable function with $E[|g(X_i)|] < \infty$. The mean of $g(\cdot)$ is balanced across inverse probability weighted instrument groups and corresponds to the unweighted population mean, i.e. \begin{align}
	E\bigg[\frac{g(X_i)Z_i}{\pi(X_i)}\bigg] = E\bigg[\frac{g(X_i)(1-Z_i)}{1-\pi(X_i)}\bigg] = E[g(X_i)]. \label{eq_BalancingGFUNCTION}
	\end{align} 
	Thus, despite allowing for correlated unobservables driving potential outcome and treatment decision, under Assumptions A.1-A.6 balancing on \textit{observed} confounders is enough to achieve causal identification for the compliers. This is due to the fact that conditional on observables the instrument is as good as randomly allocated. Thus, any imbalances in unobservables that would introduce a bias when comparing means between units from different treatment levels for always-takers, never-takers, and compliers combined do not matter for the (unidentified) population of compliers as for the latter we have that treatment is chosen according to the instrument, i.e.~$D_i(z) = z$. The remaining differences due to varying treatment selection probabilities is then accounted for by the denominator in \eqref{eq_LATEisRATIO1} that identifies the share of compliers. Thus, balancing for compliers boils down to balancing two instrument assignment groups twice for different outcomes. In its mechanic, each step corresponds to balancing different treatment groups via inverse propensity score weighting under the more restrictive selection on observables assumptions for identification of general average treatment effects \citep{imai2014covariate,zhao2019covariate}.
	
	While population balance \eqref{eq_BalancingGFUNCTION} is certainly present when true instrument propensities are used, note that when using true scores or when choosing a model for $\pi(X_i)$ in finite samples, the ultimate goal is to impose balance on conditional means, not to necessarily choose a model that best predicts $\pi(X_i)$ or $Z_i$ in a given sample according to standard loss functions such as entropy/likelihood, accuracy or mean squared error. From a population perspective or asymptotically these goals are usually aligned. For example, a correctly specified instrument propensity score obtained via maximum likelihood will eventually converge to the maximizer of the population likelihood that is the true instrument propensity score. In finite samples, however, balance is not guaranteed and hence any differences between the weighted distributions can still substantially compromise estimation and causal inference. Thus, to reduce bias choosing an estimator that exploits \eqref{eq_BalancingGFUNCTION} directly should be beneficial. 

	\section{The Balancing Estimator} \label{sec_BalEstimation}
	To impose balancing for the functions of choice we propose to use a tailored loss approach that explicitly exploits sample equivalents of \eqref{eq_BalancingGFUNCTION} for estimation of the LATE. For selection on observables, this in spirit of the tailored loss framework by \cite{zhao2019covariate} and the covariate balancing propensity score method by \cite{imai2014covariate}. In particular, we would like to estimate instrument propensity scores that yield exact balance in finite samples for a vector-valued function $\phi:\mathcal{X}\rightarrow \mathbb{R}^r$ with $r < n$. This amounts to adapting the loss function and to choosing a logistic link for the instrument propensity score with regressors $\phi(X_i) = (\phi_1(X_i),\dots,\phi_r(X_i))'$. In particular, for some $\theta \in \mathbb{R}^r$ let $L(\phi(X_i)'\theta) = 1/(1+\exp(-\phi(X_i)'\theta))$ denote the standard logistic cumulative distribution function at index $\phi(X_i)'\theta$.  The balancing estimator for $\theta$ is obtained via maximizing the following tailored (negative) loss function: \begin{align}
	\hat{\theta} &= \arg\underset{\theta}{\max}\ \frac{1}{n}\sum_{i=1}^{n}S(Z_i,X_i,\theta), \label{eq_BalancingMAXPROB1} \\
	S(Z_i,X_i,\theta) &= (2Z_i - 1)\ln\bigg(\frac{L(\phi(X_i)'\theta)}{1-L(\phi(X_i)'\theta)}\bigg) \notag \\&\quad - (Z_i - L(\phi(X_i)'\theta))\bigg(\frac{1}{L(\phi(X_i)'\theta)} - \frac{1}{1-L(\phi(X_i)'\theta)}\bigg).
	\end{align}
	This function is globally concave and has the same population maximizer as the maximum likelihood estimator for an equivalent logit model. 
	Choosing the logistic link has the advantage of imposing exact balance in finite samples as can be seen from the first order condition of the maximization problem. Setting the first derivative of \eqref{eq_BalancingMAXPROB1} equal to zero at $\theta = \hat{\theta}$ yields
	\begin{align}
	\frac{1}{n}\sum_{i=1}^{n}\frac{Z_i}{\hat{\pi}(X_i)}\phi(X_i) - \frac{1}{n}\sum_{i=1}^{n}\frac{1-Z_i}{1-\hat{\pi}(X_i)}\phi(X_i) = 0 \label{eq_BalancingFOC1}
	\end{align}
	with $\hat{\pi}(X_i) = L(\phi(X_i)'\hat{\theta})$. Thus, the tailored loss with logistic link chooses the propensity score model such that balance holds exactly for the empirical counterparts of \eqref{eq_BalancingGFUNCTION}. In principle, other link functions or balancing approaches are possible as well, see e.g.~\cite{imai2014covariate}. However, the logistic link yields a transparent connection between the choice of balancing functions and the choice of regressors in the instrument propensity score model. As problem \eqref{eq_BalancingMAXPROB1} can be understood as a just-identified moment problem, standard model diagnostics or significance tests can be applied, see also \cite{imai2014covariate} for a GMM version of this argument for selection on observables. The balanced IPW estimator for the LATE is then given by \begin{align}
	\hat{\tau}_{LATE} = \frac{\hat{\Delta}}{\hat{\Gamma}}, \label{eq_LATEbalaIPW}
	\end{align}
	with \begin{align}
	\hat{\Delta} &= \frac{1}{n}\sum_{i=1}^{n}\frac{Z_iY_i}{\hat{\pi}(X_i)} - \frac{1}{n}\sum_{i=1}^{n}\frac{(1-Z_i)Y_i}{1-\hat{\pi}(X_i)}, \\
	\hat{\Gamma} &= \frac{1}{n}\sum_{i=1}^{n}\frac{Z_iD_i}{\hat{\pi}(X_i)} - \frac{1}{n}\sum_{i=1}^{n}\frac{(1-Z_i)D_i}{1-\hat{\pi}(X_i)}. 
	\end{align}
	A particular feature that is unique to the balanced LATE approach is that if $\phi(X_i)$ contains an intercept, then the LATE estimator in \eqref{eq_LATEbalaIPW} is automatically weight-normalized, i.e.~numerically identical to its ``IPWII'' version that uses weights of the form $Z_i/\hat{\pi}(X_i)/[n^{-1}\sum_{i=1}^{n}Z_i/\hat{\pi}(X_i)]$ and $(1-Z_i)/(1-\hat{\pi}(X_i))/[n^{-1}\sum_{i=1}^{n}(1-Z_i)/(1-\hat{\pi}(X_i))]$. This follows from the ratio form of $\hat{\tau}_{LATE}$ together with \eqref{eq_BalancingFOC1} for $\phi(X_i) = c$ for any $c\neq 0$. It is \textbf{not} the case that balanced inverse probability weights themselves are normalized to unity, i.e.~in general $n^{-1}\sum_{i=1}^{n}Z_i/\hat{\pi}(X_i) \neq 1$ and equivalently for the complementary weights. Under selection on observables, there is clear evidence that weight-normalized versions of inverse probability weighting estimators generally outperform their unweighted counterparts due to a reduction in variance \citep{busso2014new,pohlmeier2016simple}. We expect these results to translate to the case of selection on unobservables. The Monte Carlo study in Section \ref{sec_BalMC1} suggests that a non-negligible part of the superior finite sample performance of the balanced IPW compared to the maximum likelihood based IPW is due to this default normalization.

	\section{Statistical Properties of the Balancing Estimator} \label{sec_BAL_properties}
	\subsection{Approximate Finite Sample Bias} \label{sec_BAL_approximateFSbias}
	A central goal of imposing the balancing constraint \eqref{eq_BalancingFOC1} is to reduce estimation bias in finite samples. Under the more restrictive selection on observables assumptions and treatment effect homogeneity, \cite{zhao2019covariate} demonstrates that a sufficiently flexible model for the conventional propensity score is enough to achieve an unbiased estimator of the treatment effect up to weight normalization. As discussed in Section \ref{sec_BalEstimation}, in contrast to the estimator by \cite{zhao2019covariate}, the balanced IPW estimator of the LATE achieves weight normalization by construction if an intercept is included into the model. Thus, it would seem straightforward to assume that the balanced LATE estimator should be equal to a ratio of two unbiased estimators if both conditional treatment effects $E[Y_i(1)-Y_i(0)|X_i]$ and conditional first stage $E[D_i(1)-D_i(0)|X_i]$ are constant independently of $X_i$. It turns out, however, that it is sufficient to have treatment effect homogeneity within the groups of compliers only and that there are no restrictions required for the conditional potential treatment to instrument responses. In particular, we obtain the following proposition:
	\begin{prop} \label{prop_BAL_unbiased1}
		For any $z\in\{0,1\}$, let $E[D_i(z)|X_i], E[D_i(z)(Y_i(1)-Y_i(0))|X_i]$ and $E[Y_i(0)|X_i]$ be in the linear span of $\{\phi_1(X_i),\phi_2(X_i),\dots,\phi_r(X_i)\}$. Under constant conditional causal effects for the compliers $E[Y_i(1)-Y_i(0)|X_i,D_i(1)>D_i(0)] = \tau_{LATE}(X_i) = \tau_{LATE}$, the inverse probability weighting estimator \eqref{eq_LATEbalaIPW} that uses balanced instrument propensity scores \eqref{eq_BalancingMAXPROB1} is a ratio of two unbiased estimators, i.e. \begin{align}
		\frac{E[\hat{\Delta}]}{E[\hat{\Gamma}]} = \tau_{LATE}. 
		\end{align}
	\end{prop}
	
	The span condition demonstrates the requirement of the instrument propensity score model to be able to contain elements that capture variation in the conditional mean of the control outcome $E[Y_i(0)|X_i]$. Moreover, the balancing components have be flexible enough to cover the relevant building blocks for the conditional mean of a potential treatment level and the conditional causal effect for a combination of different units. These conditions can be reformulated by using the monotonicity assumption. For the case with $z=0$, it reduces to the assumption that $E[D_i(0)|X_i]$ and the product between $E[D_i(0)|X_i]$ and the conditional causal effect for the always-takers $\tau_{AT}(X_i)$ have to be contained in the linear span as \begin{align}
	E[&D_i(0)(Y_i(1)-Y_i(0))|X_i] \notag \\ &= E[Y_i(1)-Y_i(0)|X_i,D_i(0)=1,D_i(1)=1]P(D_i(0)=1,D_i(1)=1|X_i) \notag  \\&\quad + E[Y_i(1)-Y_i(0)|X_i,D_i(0)=1,D_i(1)=0]P(D_i(0)=1,D_i(1)=0|X_i) \notag \\
	&= E[Y_i(1)-Y_i(0)|X_i,D_i(0)=1,D_i(1)=1]P(D_i(0)=1|X_i)\notag  \\
	&= \tau_{AT}(X_i)E[D_i(0)|X_i] 
	\end{align}
	with the second equality following from monotonicity. For the case of $z=1$, it is required that the linear combination of probability weighted versions of the causal effects of compliers and always-takers are captured since \begin{align}
	E[&D_i(1)(Y_i(1)-Y_i(0))|X_i] \notag  \\ &= E[Y_i(1)-Y_i(0)|X_i,D_i(1)>D_i(0)]P(D_i(1)>D_i(0)|X_i) \notag  \\&\quad + E[Y_i(1)-Y_i(0)|X_i,D_i(1)=1,D_i(0)=1]P(D_i(1)=1,D_i(0)=1|X_i) \notag  \\
	&= \tau_{LATE} E[D_i(1)-D_i(0)|X_i] + \tau_{AT}(X_i)E[D_i(0)|X_i]
	\end{align}
	with the second equality following from constant complier causal effects and monotonicity. Thus, the conditions in Proposition \ref{prop_BAL_unbiased1} effectively demand a special type of flexibility regarding which transformations of regressors $\phi(X_i)$ should be contained in the balancing constraints for the instrument propensity score. Depending on the application at hand, these conditions can be rather restrictive. However, note that the derivations required for Proposition \ref{prop_BAL_unbiased1} reveal that it is actually not necessary that the instrument scores used in the denominator have the same degree of flexibility as the ones in the numerator. In particular, for denominator scores only a condition for $E[D_i(z)|X_i]$ for a $z\in\{0,1\}$ is required. This can be exploited by choosing different instrument propensity score models for numerator and denominator. We return to this point in Section \ref{sec_BAL_higherorder1}. 
	
	The robustness property in Proposition \ref{prop_BAL_unbiased1} is best understood in comparison with the two-stage least squares (2SLS) estimator. In general, if the first stage for the endogenous treatment variable is fully saturated, the model behind 2SLS identifies a weighted version of conditional LATEs with weights being proportional to the conditional variances of the first stages $V[E[D_i|X_i,Z_i]|X_i]$, see \cite{angrist1995two}. Under homogeneous treatment effects, however, this corresponds to the LATE. Morever, the parameters from the reduced forms for both outcome and endogenous variable can be estimated without bias. Thus, 2SLS is a ratio of two unbiased estimators with the ratio of the expectations being equal to the LATE. Proposition \ref{prop_BAL_unbiased1} states that the balanced IPW estimator has an equivalent property under comparable assumptions. However, instead of having a flexible model for the treatment choice $E[D_i|X_i,Z_i]$, the flexibility is incorporated through the choice of the variables and transformations $\phi(X_i)$ in the model for instrument propensity score $E[Z_i|X_i]$. For IPW estimators of the LATE, the finite sample bias property in Proposition \ref{prop_BAL_unbiased1} is unique to the balanced instrument propensities and generally does not apply to any other estimation approach such as maximum likelihood or even true instrument propensity scores. We also expect this property to be beneficial in finite samples if there are moderate deviations from homogeneous causal effects for the compliers. In general, the ratio of the expectations is only an approximation to the expectation of the ratio in finite samples. 
	Therefore, the actual finite sample bias for has to be further investigated. We return to this point in Section \ref{sec_BalMC1}.

	\subsection{Duality and Variance Reduction} \label{sec_BAL_CondVarDuality}
	In this section, we show that in finite samples the balancing approach favors moderate instrument propensity scores in the sense of being close to one half. As instrument propensity scores are inversely related to the (conditional) variance of the IPW estimator for the LATE, there are potential gains in terms of point estimation risk compared to using likelihood scores. For moderate inverse probability weights, the dual problem of maximizing the (negative) tailored loss in \eqref{eq_BalancingMAXPROB1} penalizes deviations from the unconditional mean in a manner that is proportional to the conditional variance of the estimator for the LATE. Let the inverse probability weights be defined as \begin{align}
	w_i = \frac{Z_i}{\hat{\pi}(X_i)} + \frac{1-Z_i}{1-\hat{\pi}(X_i)}
	\end{align} and denote $W = (W_1,\dots,W_n)$ with $W_i = (X_i',Z_i)$ for $i=1,\dots,n$. Conditional on $W$, the balanced instrument propensity scores are known. Thus, a second order Taylor expansion of the variance of the balanced LATE \eqref{eq_LATEbalaIPW} conditional on $W$ yields \begin{align}
	nV\bigg[\frac{\hat{\Delta}}{\hat\Gamma}\bigg|W\bigg] &\approx \frac{1}{n}\sum_{i=1}^{n}w_i^2 a(X_i,Z_i) 
	\end{align}
	with
	\begin{align} \label{eq_BAL_aweightVAR1}
	a(X_i,Z_i) &= \bigg(V[Y_i|W_i]E[\hat{\Gamma}|W]^2 - 2Cov(Y_i,D_i|W_i)E[\hat{\Delta}|W]E[\hat{\Gamma}|W] + \notag \\ & \ \quad \ E[D_i|W_i]E[\hat{\Delta}|W]^2\bigg){E[\hat{\Gamma}|W]^{-4}}
	\end{align}  
	which is strictly greater than zero. Hence, approximately the conditional variance is bounded from above by \begin{align}
	nV\bigg[\frac{\hat{\Delta}}{\hat\Gamma}\bigg|W\bigg] \leq \underset{x\in\mathcal{X}, z \in\{0,1\}}{\sup}a(x,z)\ \frac{1}{n}\sum_{i=1}^{n}w_i^2
	\end{align}
	which is proportional to the average of the squared inverse probability weights. The conditional variance is directly proportional to the latter, i.e.~$nV[{\hat{\Delta}}/{\hat\Gamma}|W] \sim n^{-1}\sum_{i=1}^{n}w_i^2$ if there is homoskedasticity within outcomes and treatment levels and constant correlation across $Y_i$ and $D_i$ conditional on $W_i$ or if the corresponding components in the numerator of \eqref{eq_BAL_aweightVAR1} are proportional such that $a(x,z) = a$ for all $x\in\mathcal{X}$ and $z\in\{0,1\}$. To approximate the effect of using a balancing estimator on the squared weights, it is insightful to study the Lagrangian dual problem of the maximization problem \eqref{eq_BalancingMAXPROB1} in terms of the weights $w_i$. If $\phi(X_i)$ contains an intercept, the dual problem is given by the following constraint optimization: \begin{align}
	\underset{w_1,\dots,w_n}{\min}\ &\quad \frac{1}{n}\sum_{i=1}^{n}(w_i - 1)\ln(w_i - 1) - w_i, \notag \\
	\text{subject to } &\quad \frac{1}{n}\sum_{i=1}^{n}(2Z_i-1)w_i\phi_m(X_i) = 0, \ \text{ for } \ m =1,\dots,r \notag \\
	&\quad w_i \geq 1 \ \text{ for } \ i=1,\dots,n, \label{eq_BAL_dual1}
	\end{align}
	see also \cite{chan2016globally} and \cite{zhao2019covariate}. While this is generally not equivalent to minimizing the sum of squared weights, consider the case of an independent instrument such as assignment to treatment in a randomized experiment with imperfect compliance. Under such a circumstance, units receive instrument levels with constant likelihood, e.g.~$P(Z_i=1) = 0.5$. The true inverse probability weights are then given by $1/P(Z_i=1) = 2$. Naturally, even under perfect randomization there can be imbalances im terms of the covariates across the two instrument groups in finite samples. Thus, balancing scores $\hat{\pi}(X_i)$ will generally differ from the true inverse probability weights. Consider a deviation from the true probability of one half to a more extreme one, i.e.~an inverse probability weight exceeding two. A Taylor series of the unconstrained loss function for a single observation in the dual problem \eqref{eq_BAL_dual1} at $w = 2$ yields 
	\begin{align} \label{eq_BAL_dualapprox1}
	(w - 1)\ln(w - 1) - w = -2 + \frac{1}{2}(w-2)^2 + O((w-3)^3)
	\end{align}
	which is a convergent series for moderate deviations, i.e.~$|w -2| < 1$. For more general deviations from moderate instrument propensities, the tailored loss behaves qualitatively similar. Expansion \eqref{eq_BAL_dualapprox1} reveals that the dual problem locally penalizes deviations from the equal weighting in a quadratic manner. As equal weighting is variance minimizing by construction\footnote{Equal weighting corresponds to the simple Wald estimator that only relies on binary reduced forms and does not use any covariates. The conditional variance of the latter is proportional to $1/(P(Z_i=1)(1-P(Z_i=1)))$ and thus always below the conditional variance of the IPW estimator that is equally proportional to $n^{-1}\sum_{i=1}^{n}w_i^2$ using true instrument propensities under homoskedasticity for the different assignment groups.}, this implies that, approximately, the balancing weights seek to minimize an upper bound for the conditional variance of the LATE estimator. They do not succeed in imposing an exactly constant weighting scheme due to the constraints in \eqref{eq_BAL_dual1} that are designed to minimize any bias from imbalances across assignment groups. 
	
	Moderate weights and balancing covariates without perfect randomization are generally opposing goals. In principal, one could also use the characterization of the conditional variance in \eqref{eq_BAL_aweightVAR1} to choose minimizing weights within a class of balancing weights as proposed in the context of selection on observables by \cite{li2016balancing}. This strategy, however, changes the definition of the underlying identified causal parameter from the LATE to a ratio of two weighted reduced form estimates which put a higher weight to individuals with instrument propensities close to one half. The resulting identified parameter does not lend itself to an intuitive causal interpretation with similar policy relevance compared to the conventional LATE. 
	
	\subsection{Nonparametric Estimation and Large Sample Properties} \label{sec_BAL_NONPAR1}
	In light of Proposition \ref{prop_BAL_unbiased1} it seems reasonable to choose a model for the instrument propensity score that eventually incorporates a large set of balancing constraints as the sample size increases. In this section we show that using a nonparametric approach within the tailored loss framework can efficiently incorporate all required information for estimation of the LATE in a semiparametric sense. It does so while still retaining the exact finite sample balancing property in \eqref{eq_BalancingFOC1}. We rely on a series approach using power series similar to the series logistic approach proposed by \cite{hirano2003efficient} and \cite{donald2014testing} that rely on a local maximum likelihood step for estimation of the (instrument) propensity scores.
	For $K> 0$, let $\phi^K(x) = (x^{\lambda(1)},\dots,x^{\lambda(K)})'$ be a vector of power functions such that $||\lambda(k)||_1 \leq ||\lambda(k+1)||_1$ for $k \in \mathbb{N}_0^r$ with $\lambda = (\lambda_1,\dots,\lambda_r)'$ nonnegative and $||\cdot||_1$ denoting the $\ell_1$-norm, i.e.~$||\lambda||_1 = \sum_{j=1}^{r}|\lambda_j|$. We assume that the series is orthogonalized with respect to a weight function such that $E[\phi^K(X_i)\phi^K(X_i)'] = I_K$. This is always possible since a logistic link function is used and thus we approximate the log-odds ratio by a linear single index $\phi^K(x)'\theta_K = \theta_K'A_k^{-1}A_k\phi^K(x)$. Therefore, one can always use basis $A_k\phi^K(x)$ for approximation instead, see also Appendix A in \cite{hirano2003efficient}. The balanced instrument propensity scores are then given by $\hat{\pi}(x) = L(\phi^K(x)'\hat{\theta}_K)$ with \begin{align}
	\hat{\theta}_K &= \arg\underset{\theta}{\max}\ \frac{1}{n}\sum_{i=1}^{n}S(Z_i,X_i,\theta), \label{eq_BalancingMAXPROB_NONPAR} \\
	S(Z_i,X_i,\theta) &= (2Z_i - 1)\ln\bigg(\frac{L(\phi^K(X_i)'\theta)}{1-L(\phi^K(X_i)'\theta)}\bigg) \notag \\&\quad - (Z_i - L(\phi^K(X_i)'\theta))\bigg(\frac{1}{L(\phi^K(X_i)'\theta)} - \frac{1}{1-L(\phi^K(X_i)'\theta)}\bigg).
	\end{align} 
	Let $C^d$ denote the space of $d$-times continuously differentiable functions. We impose the following regularity and smoothness assumptions: \textit{
		\begin{enumerate}
			\item[B.1)] $X_i$ are $r$-dimensional random variables compactly supported on $\mathcal{X}$ with absolutely continuous density $f(x)$ in $C^2$ and bounded away from zero.
			\item[B.2)] $E[Y_i|X_i,Z_i=z] = m_z(X_i)$ and $E[D_i|X_i,Z_i=z] = \mu_z(X_i)$ are in $C^1$.
			\item[B.3)] $\pi(X_i)$ are in $C^q$ with $q\geq 7r$.
			\item[B.4)] All second moments of potential outcome levels exist and are finite.
			\item[B.5)] $K = O(n^{v})$ with $1/(4(q/r-1))< v < 1/9$.
		\end{enumerate}
	}
	The assumptions are standard in the literature, see  \cite{hirano2003efficient} and \cite{donald2014testing} or \cite{li2009efficient} and \cite{donald2014inverse,donald2014testing} for possible restrictions on the series terms and adaptations to discrete covariates and kernel methods. Assumption B.1 assures a uniform approximation of any continuous function of the covariates, in particular conditional means such as the instrument propensity score. Assumptions B.2 and B.3 are smoothness conditions on the observed outcome and treatment status conditional on covariates and instrument level and on the instrument propensity score. The higher the dimensionality of the covariates, the more smoothness conditions are required. B.4 is a regularity condition that is necessary to assure a finite asymptotic variance for the LATE estimator. Assumption B.5 controls the rate at which the order of basis functions is allowed to grow with increasing sample size depending on the degree of smoothness.  We obtain the following theorem: 	\begin{theorem}[Efficient Balancing] \label{theorem_BAL_efficient1}
		Under Assumptions A.1-A.6, B.1-B.5, and instrument propensity scores estimated according to \eqref{eq_BalancingMAXPROB_NONPAR}, the inverse probability weighting estimator for the LATE 
		\begin{enumerate}
			\item is asymptotically normal $$\sqrt{n}(\hat{\tau}_{LATE} - \tau_{LATE}) \overset{d}{\rightarrow} \mathcal{N}(0,V)$$
			\item and reaches the semiparametric efficiency bound \begin{align*}
			V &= \frac{1}{\Gamma^2}\bigg(E[(m_1(X_i) - m_0(X_i) - \tau_{LATE}\mu_1(X_i) + \tau_{LATE}\mu_0(X_i))^2] \\ &\quad + \sum_{z=0,1} E\bigg[\frac{\sigma_{Y_z}^2(X_i) - 2\tau_{LATE}\sigma^2_{Y_zD_z}(X_i) + \tau_{LATE}^2\sigma_{D_z}^2(X_i)}{P(Z_i=z|X_i)}\bigg] \bigg)	
			\end{align*}
			with $\sigma_{Y_z}^2(X_i) = V[Y_i|X_i,Z_i=z]$, $\sigma_{D_z}^2(X_i) = V[D_i|X_i,Z_i=z]$, and $\sigma_{Y_zD_z}^2(X_i) = Cov[Y_i,D_i|X_i,Z_i=z]$ for any $z\in\{0,1\}$.
		\end{enumerate}
	\end{theorem} 
	Theorem \ref{theorem_BAL_efficient1} shows that the inverse probability weighting estimator using a sufficiently flexible nonparametric model for the instrument propensity score that imposes empirical balancing constraints efficiently incorporates all information available for estimation of the LATE and is consistent. This is in line with the insights from the bias characterizations in Section \ref{sec_BAL_approximateFSbias}. There, the instrument propensity scores yield first-order unbiased estimates if they incorporate the structure of the potential treatment effects and potential treatment levels sufficiently. If these conditional mean functions are continuous, then a series approximation will eventually contain all the components necessary to uniformly approximate them on compact sets. The proof of Theorem \ref{theorem_BAL_efficient1} mainly relies on the global concavity of the tailored loss function together with the strong instrument overlap assumption. Theorem \ref{theorem_BAL_efficient1} also implies that in large samples there is no qualitative difference between tailored instrument propensity scores and standard nonparametric approaches. However, the balancing approach additionally guarantees finite sample balance. The efficiency bound for the LATE has originally been derived by \cite{frolich2007nonparametric}, see also \cite{hong2010semiparametric}. The asymptotic variance $V$ can be consistently estimated using the series approach as in \cite{hirano2003efficient} and \cite{donald2014testing} by replacing the population quantities with sample estimates, see Appendix \ref{sec_BAL_AsymptoticVar} for more details.
	
	\section{Extensions} \label{sec_BAL_extensions}
	\subsection{Using Higher-order Information to Improve Balance} \label{sec_BAL_higherorder1}
	The results on nonparametric estimation in Section \ref{sec_BAL_NONPAR1} make a strong case for eventual flexibility of the balancing constraints used for estimation of instrument propensity scores. In large samples, however, the exact choice or order of inclusion of transformations of regressors is not of primary concern. Thus, the question remains, which empirical means should be prioritized for balancing in finite samples, i.e.~how to pick $\phi(X_i)$? From a model selection perspective, choosing informative balancing variables first will be beneficial if they do not render the estimation step infeasible or instable due to e.g.~multicollinearity. It might even be useful to exploit a set of generated regressors whose balancing is beneficial for precise estimation of the treatment effect. Looking at the LATE from an applied perspective, it is often reasonable to assume a hierarchy in terms of the knowledge about the different steps of the underlying causal mechanism. This hierarchy usually goes in ascending order from a) assignment over b) treatment choice to c) outcome process. Consider the example of evaluating the causal impact of a job search training program on future earnings for the unemployed offered by an employment agency. In this case, the decision to assign units to a program might be based on a set of observable characteristics such as age, employment history, and qualifications available to the agent responsible for assignment. The instrument assignment on level a) is conditionally independent if all other variables that affect the assignment decision are independent of the potential earnings and potential treatment levels. On hierarchy level b), the unemployed units choose whether to comply with the assignment. If enough information about their trade-offs and restrictions are available, we can model the choice problem and its constraints guided by economic theory. The outcome process determining earnings, however, is likely generated as a consequence of a complicated searching and matching process on the labor market under additional constraints. Thus, relying only on data from the employment agency might not be enough to really inform a model for  hierarchy level c). 
	
	In light of Proposition \ref{prop_BAL_unbiased1}, balancing the following quantities would in principle be desirable: \begin{enumerate}
		\item[(i)] $E[Y_i(0)|X_i], E[D_i(z)(Y_i(1)-Y_i(0))]$ and
		\item[(ii)] $E[D_i(z)|X_i]$ 
	\end{enumerate}    
	for any $z \in \{0,1\}$. (i) requires outcome data to generate e.g.~model-based quantities. Hence, using them for empirical balancing constraints operates on the highest hierarchy level c) and goes against the design arguments outlined by \cite{rubin2007design}.  (ii) on the other hand is a better candidate as it only concerns the treatment choice step. For simplicity assume that $X_i$ is discrete. From the results of \cite{vytlacil2002independence} it follows that Assumptions A.1-A.4 (conditional on $X_i$) imply the existence a nonparametric single index model that rationalizes the identical choices as the LATE framework conditional on each $x \in  \mathcal{X}$ and vice versa, see also \cite{kline2019heckits} for estimation and numerical equivalence. Thus, it motivates a fully nonparametric model for the first stage. In particular, one obtains \begin{align}
	D_i = \mathbbm{1}(\mu(X_i,Z_i) > v_i) \label{eq_BAL_STRUCTURALmodel1}
	\end{align}
	with $v_i \ \indep \ Z_i|X_i$, $\mu(x,z)$ non-degenerate conditional on $x$, and $F_{v|x}(v)$ continuous. Without any further restrictions and $X_i$ discrete, this would correspond to the first stage of a fully saturated instrumental variables approach. The estimates for selection probabilities $E[D_i(z)|X_i]$ are obtained as the estimates for $F_{v|x}(\mu(X_i,z))$. They can then be included into the vector of transformations of regressors $\phi(X_i)$ to balance the empirical counterpart of $E[D_i(z)|X_i]$ in finite samples.\footnote{For a regressor with many categories or multiple discrete regressors, using additional smoothing methods as in \cite{ouyang2009nonparametric} or \cite{heiler2018shrinkage} for estimation of \eqref{eq_BAL_STRUCTURALmodel1} might be desirable.} The proof of Proposition \ref{prop_BAL_unbiased1} (see Appendix \ref{app_BAL_proofBIASratio}) reveals that this strategy is enough for the denominator to fulfill its role in having an estimator for the LATE under conditional independence that is given by the ratio of two unbiased estimators. For the numerator, however, information about the outcome process as in (i) would be required. If one is willing to impose a model for the potential outcomes, then from a bias perspective it would be sufficient to include them into the balancing constraints for the instrument propensity scores that enter the numerator only. Thus one can in principal operate with two different instrument propensity scores for numerator and denominator that differ by using balancing constraints for the empirical counterparts of \textbf{either} $E[D_i(z)|X_i]$ (denominator) \textbf{or} $E[Y_i(0)|X_i]$ and $E[D_i(z)(Y_i(1)-Y_i(0))|X_i]$ (numerator). Note that under constant treatment effects for some $z\in\{0,1\}$, i.e.~for always-takers and compliers or never-takers and compliers we have that $E[D_i(z)(Y_i(1)-Y_i(0))|X_i] = E[D_i(z)|X_i]\tau$. Thus, the second component in (i) is balanced by simply including $E[D_i(z)|X_i]$ into the instrument propensity score model. Therefore, using a single instrument propensity score for both numerator and denominator that only additionally incorporates $E[D_i(z)|X_i]$ for balancing seems like a reasonable middle-ground between the fully agnostic approach and imposing a lot of structure on the outcome process.
	
	If a parametric model for $\mu(X_i,Z_i)$ is used, there are two additional aspects to consider. First, statistical inference has to be adjusted to the presence of the generated regressors.\footnote{This can be done by a standard asymptotic expansion that captures the additional parametric uncertainty in estimating $\mu(X_i,Z_i)$ or by using a bootstrap approach.} Second, there are two potentially highly correlated choices for inclusion into the balancing constraints, $E[D_i(1)|X_i]$ and $E[D_i(0)|X_i]$. In general, one of these components is enough for approximate unbiasedness. If $\phi(X_i)$ also contains other transformations of regressors, it seems reasonable to include the balancing constraint that adds more information. Including both can quickly lead to multicollinearity problems, in particular if variation in the instrument only has a mild effect on the conditional probability of choosing a certain treatment, i.e.~if the expected share of compliers is not very large. This phenomenon and general finite sample performance of the different strategies are further investigated in Section \ref{sec_BalMC1}.

	\subsection{High-dimensional Models and Regularization} \label{sec_BAL_regularization1}
	If the dimensionality of the transformation of regressors is large relative to the sample size, i.e.~if $r > n$, solving for the parameters of the balancing constraints in \eqref{eq_BalancingFOC1} is generally infeasible. However, with the cost of imposing some bias, balancing can be relaxed to tolerance level which makes estimation feasible. There is a growing literature on approximate balancing approaches in the context of estimation of treatment effects under selection on observables using standard propensity scores. \cite{zubizarreta2015stable} proposes to minimize the squared $\ell_2$-norm of the balancing weights subject to an empirical balancing constraint. \cite{athey2018approximate} combine approximate balancing using a supremum-type constraint with a high-dimensional linear model for the outcome under mild sparsity conditions for bias correction. \cite{zhao2019covariate} suggests the use of regularized generalized linear models, reproducing kernel Hilbert spaces, and boosted trees as penalization strategies for general tailored loss functions. There are no intrinsic differences to the case of the LATE using balanced instrument propensity scores. We briefly outline the general principle of approximate balancing adapted to the tailored loss considered in this paper with a focus on commonly used parameter norm constraints. Let the transformation of regressors $\phi(X_i)$ be standardized. Consider the penalized optimization problem \begin{align}
	\hat{\theta}_{\lambda} &= \arg\underset{\theta}{\max}\ \frac{1}{n}\sum_{i=1}^{n}S(Z_i,X_i,\theta) - \lambda J(\theta) \label{eq_BalancingMAXPROB1_REGULARIZED1} 
	\end{align}  
	with $J:\mathbb{R}^r\rightarrow{\mathbb{R}}$ being a (convex) penalty function. Typical  choices are lasso $J(\theta) = ||\theta||_1$, ridge regression $J(\theta) = ||\theta||_2^2/2$, combinations thereof (elastic net), or any $\ell_a$-norm $J(\theta) = ||\theta||_a^a/a$ for $a \geq 1$. Convexity of $J(\cdot)$ assures that \eqref{eq_BalancingMAXPROB1_REGULARIZED1} is equivalent to the following dual problem \citep{zhao2019covariate}:
	\begin{align}
	\underset{w_1,\dots,w_n}{\min}\ &\quad \frac{1}{n}\sum_{i=1}^{n}(w_i - 1)\ln(w_i - 1) - w_i, \notag \\
	\text{subject to } &\quad \bigg|\frac{1}{n}\sum_{i=1}^{n}(2Z_i-1)w_i\phi_m(X_i)\bigg| \leq \lambda |({\theta}_{\lambda})_m|^{a-1}, \ \text{ for } \ m =1,\dots,r \notag \\
	&\quad w_i \geq 1 \ \text{ for } \ i=1,\dots,n, \label{eq_BAL_dualREGULARIZED1}
	\end{align}
	with weights $w_i$ depending on $\theta_{\lambda}$ as before. Thus, penalization relaxes the empirical balancing constraint from exact balance to tolerance level depending on the penalty of choice. In the simple case of a lasso type penalty, i.e.~$a=1$, the empirical balancing is relaxed to tolerance level $\lambda$. For $\lambda = 0$, \eqref{eq_BalancingMAXPROB1_REGULARIZED1} is equivalent to exact balancing while for $\lambda \rightarrow \infty$ all weights are set to a constant by the same argument as for the non-regularized dual problem in Section \ref{sec_BAL_CondVarDuality}. In case of the latter, the variance of the estimator is minimized as in Section \ref{sec_BAL_CondVarDuality}. However, as balancing constraints are imposed to decrease bias, see Section \ref{sec_BalEstimation} and Section \ref{sec_BAL_approximateFSbias}, a larger amount of regularization will introduce some bias into the estimation of the LATE. To solve this trade-off, \cite{zhao2019covariate} proposes to select $\lambda$ via cross-validation using the covariate imbalance in the validation set for tuning and to compare standardized differences along different choices of $\lambda$. 
	
	\section{Monte Carlo Study} \label{sec_BalMC1}
	\subsection{Design}
	In this section, we compare the finite sample performance of the balancing estimator and some of the proposed extensions to standard estimation approaches from the literature. In particular, the question arises whether balancing approaches can outperform conventional methods in terms of point estimation risk and if so, what are the driving forces behind it. In Sections \ref{sec_BAL_approximateFSbias} and \ref{sec_BAL_CondVarDuality} we provide some reasoning why reductions in points estimation risk from balancing could be due to both channels, bias and variance. We evaluate the impact of using higher-order information from the treatment selection step as proposed in Section \ref{sec_BAL_higherorder1} and disentangle it from the additional estimation noise and the effects of simple weight normalization. We consider multiple designs that highlight different features of the balancing approaches compared to conventional methods. Table \ref{tab_BAL_MC_genRoy1} contains the basic Monte Carlo design in spirit of a generalized Roy model:\footnote{See \cite{heckman2005structural} for a simulation of a simple generalized Roy model using a continuous instrument.} 
	
	\begin{table}[!h]
		\centering \caption{Monte Carlo Study: Generalized Roy Model}  \label{tab_BAL_MC_genRoy1}
		\begin{minipage}{0.4\textwidth} 
			\begin{align*}
			X_i &\sim Uniform(0,1)   \\
			Z_i&= \mathbbm{1}(u_i < \pi(X_i))\\
			D_i(z) &= \mathbbm{1}(\mu_d(X_i,z) > v_i) \\
			Y_i(1) &= \mu_{y_1}(X_i) + \varepsilon_i(1)\\
			Y_i(0) &= \varepsilon_i(0)
			\end{align*}	
		\end{minipage} 
		\begin{minipage}{0.4\textwidth}  \begin{align*}
			\qquad \pi(X_i)&= 1/(1+\exp(-(2X_i-1)\theta_0))\\
			u_i  &\sim Uniform(0,1)  \\ 
			\begin{pmatrix}
			\varepsilon_i(1) \\ \varepsilon_i(0) \\ v_i
			\end{pmatrix} &\sim \mathcal{N} \begin{pmatrix}
			\begin{bmatrix}
			0 \\ 0 \\ 0
			\end{bmatrix},\begin{bmatrix}
			1 \ 0 \ \rho \\ 0 \ 1 \ 0 \\ \rho \ 0 \ 1 
			\end{bmatrix} 
			\end{pmatrix} \end{align*}
		\end{minipage}
	\end{table}
	with $\rho = 0.5$, $\theta_0 = \ln((1-\delta)/\delta)$ and $\mu_{d}(\cdot)$ and $\mu_{y_1}(\cdot)$ see below. 
	The parameter $\theta_0$ controls the degree of overlap based on the desired bounds $(\delta,1-\delta)$ for the instrument propensity score. $\rho \neq 0$ allows for selection on unobservables. We assume a nonzero correlation between unobservables driving the potential treatment outcome and treatment selection only to simplify analysis. This is without loss of generality as in the generalized Roy model under normality the functional form of the treatment effect is not affected by this choice up to a scaling factor depending on $\rho$.\footnote{The contribution of the unobservables to the conditional LATE is of the form $\rho f(X_i)$. If we assume that $Cov(\varepsilon_i(0),v_i) = \rho_0$, then the control function simply changes to $(\rho - \rho_0) f(X_i)$. As these correlations can have different signs, one could in principle increase the overall contribution of selection from unobservables to the conditional treatment effect for a given share of compliers.} Moreover, the model allows for additive separable contributions of observables and unobservables to the overall treatment effect which simplifies calculations. In particular, the LATE parameter for the model in Table \ref{tab_BAL_MC_genRoy1} is given by \begin{align}
	\tau_{LATE} = \frac{E[\mu_y(X_i)(F(\mu_d(X_i,1))-F(\mu_d(X_i,0)))]}{E[F(\mu_d(X_i,1))-F(\mu_d(X_i,0))]} + \rho \frac{E[f(\mu_d(X_i,0))-f(\mu_d(X_i,1))]}{E[F(\mu_d(X_i,1))-F(\mu_d(X_i,0))]} \label{eq_BAL_MC_LATEtrue1}
	\end{align}
	with $F(\cdot)$ and $f(\cdot)$ being the cumulative distribution function and the density function of the univariate standard normal distribution respectively. Table \ref{tab_BAL_MC_designsALL1} contains the different design specifications for the treatment selection step and the potential treatment outcome.
	
	\begin{table}[!h]
		\centering \caption{Monte Carlo Study: Designs} \label{tab_BAL_MC_designsALL1}
		\begin{tabular}{lccc} \hline \hline 
			& Design A & Design B & Design C \\ \hline
			$\mu_d(x,z)$ 	& $4z$ & $-1 + 2x + 2.122z$ &$-1 + 2x + 2.122z$ \\
			$\mu_{y_1}(x)$  	&$0.3989$ & $0.3989$ & $9(x+3)^2$ \\ \hline 
		\end{tabular}
	\end{table}
	
	The Roy model together with the functional form assumptions assure that none of the identification conditions for the LATE are violated. In particular, the linear index for the treatment choice with the given parameters imposes monotonicity and a share of compliers $E[D_i(1)-D_i(0)] = 0.5$ for all designs. 
	In design A and design B, heterogeneity in causal effects is achieved through the correlation $\rho$ of the unobservable variables $\varepsilon_i(1)$ and $v_i$. For design C, there is heterogeneity stemming from both the direct contribution of observables to the potential outcome one and from the correlated unobservable variables. 
	Design A represents the special case of a fully independent instrument as the observed confounder does not affect the treatment choice. It also corresponds to a controlled randomized experiment with close to one-sided noncompliance as $P(D_i=1|Z_i=1) > 0.9999$. The choice of the homogeneous mean function for the outcome assures that the unconditional LATE in \eqref{eq_BAL_MC_LATEtrue1} is composed of two equally sized components, i.e.~the overall contributions of the potential outcome mean and the ``control function'' part are identical. For this design, a simple Wald estimator would yield precise estimates for the LATE. Design B is a conditionally independent design with a homogeneous potential outcome mean function. The parameters are chosen to be favorable towards linear IV with exogenous covariates $X_i$ and instrument $Z_i$. Design C is a conditionally independent design with a nonlinear potential outcome mean. In this design, the semiparametric approaches that do not require specification of an outcome model still yield asymptotically unbiased results. The precise functional form of the potential outcome mean is not crucial. In general, other nonlinear mean functions with sufficient heterogeneity will produce qualitatively similar results. 
	
	\begin{table}[!h] \centering \caption{Monte Carlo Study: Estimation Methods} \label{tab_BAL_MC_methods1}
		\begin{tabularx}{\textwidth}{l|X} \hline \hline 
			Name & Description \\ \hline 
			IV & Instrumental variables estimator using binary instrument $Z_i$ and additional control $X_i$. \\
			MLE & IPW estimator for the LATE using correctly specified maximum likelihood instrument propensity scores. \\
			MLE(2) & MLE with weight normalization (``IPWII''-type). \\
			B($X$) &Balanced IPW with $\phi(X_i) = (1\ X_i)'$. \\
			B($D$) &Balanced IPW with $\phi(X_i) = (1\ E[D_i(0)|X_i])'$ (infeasible). \\
			B($D,X$) &Balanced IPW with $\phi(X_i) = (1\ E[D_i(0)|X_i]\ X_i)'$ (infeasible).\\
			B($\hat{D}$) &Balanced IPW with $\phi(X_i) = (1\ \hat{E}[D_i(0)|X_i])'$ with $\hat{E}[D_i(0)|X_i]$ obtained from binary maximum likelihood estimation using a correctly specified probit model. \\
			B($\hat{D}_m$) & Balanced IPW with $\phi(X_i) = (1\ \hat{E}[D_i(0)|X_i])'$ with $\hat{E}[D_i(0)|X_i]$ obtained from binary maximum likelihood estimation using a misspecified logit model. \\ \hline 
		\end{tabularx}
	\end{table}
	Table \ref{tab_BAL_MC_methods1} contains all estimation approaches used in the Monte Carlo study. IV denotes the standard instrumental variables estimator that differs from the Wald estimator by additionally including $X_i$ as exogenous variable into the first-stage and the outcome equation. MLE and MLE(2) are the IPW estimators for the LATE using likelihood instrument propensities with (MLE(2)) and without (MLE) normalization of the inverse probability weights. B($X$) is the basic balancing estimator using the same regressors as the maximum likelihood approaches. As it contains an intercept, it is automatically weight normalized. B($D$) and B($D,X$) try to exploit the conditions for approximate unbiasedness in Proposition \ref{prop_BAL_unbiased1} by using $E[D_i(0)|X_i]$ as additional (or only) variable for balancing. Theoretically, both should be able to reduce the bias of the components of the LATE. Despite their general infeasibility, they are included to see the potential gains from including additional information without the statistical noise from estimation of the generated regressor $E[D_i(0)|X_i]$. B($\hat{D}$) is similar to B($D$) but with estimated quantities, i.e.~it uses a correctly specified model for $E[D_i(0)|X_i]$ by extracting the probability predictions from a probit model that uses $Z_i$ and $X_i$ as regressors at $Z_i = 0$. B($\hat{D}_m$) is similar to B($\hat{D}$) but uses a misspecified logit model instead of the true probit model. Note that, when incorrect specifications for the treatment selection are used as additional balancing constraints, this does not affect the asymptotic validity of the balancing IPW estimator for the LATE as even a misspecified quantity as a function of $X_i$ still leads to a valid balancing constraint but not necessarily one that minimizes bias in finite samples in the sense of Proposition \ref{prop_BAL_unbiased1}.  
	
	\subsection{Results}
	Tables \ref{tab_BAL_MC_DesignA_1}, \ref{tab_BAL_MC_DesignB_1}, and \ref{tab_BAL_MC_DesignC_1} contain the mean squared errors and absolute biases for designs A, B, and C and all estimation approaches from Table \ref{tab_BAL_MC_methods1}. Mean squared errors are all normalized by the MSE of the linear instrumental variables estimator. Results are obtained from an experiment using 20000 Monte Carlo replications with overlap parameters $\delta =0.01, 0.02, 0.05$ and sample sizes $n=500, 1000$ yielding an expected number of $250$ and $500$ complier units respectively. 
	
	\begin{table} \scriptsize
		\centering \caption{Design A: Relative Mean Squared Errors and Absolute Biases} \label{tab_BAL_MC_DesignA_1}
		\begin{tabularx}{\textwidth}{ll *{8}{Y}}
			\hline \hline \\
			&		&	IV	&	MLE	&	MLE(2)	&	B($X$)	&	B($D$)	&	B($D$,$X$)	&	B($\hat{D}$)	&	B($\hat{D}_m$)	\\ \hline 
			\\[-1.6ex] \multicolumn{2}{c}{$\delta = 0.01$} &&&&&&&& \\	\cmidrule(lr){1-2}
			$n = 500$	&	MSE	&	1.0000	&	336.0527	&	2.7960	&	2.7641	&	0.4411	&	2.7641	&	2.7330	&	2.7278	\\
			&	|BIAS|	&	0.0079	&	0.0574	&	0.0240	&	0.0237	&	0.0012	&	0.0237	&	0.0229	&	0.0228	\\
			&		&		&		&		&		&		&		&		&		\\
			$n = 1000$	&	MSE	&	1.0000	&	15.9247	&	2.6831	&	2.6903	&	0.4450	&	2.6903	&	2.6665	&	2.6637	\\
			&	|BIAS|	&	0.0044	&	0.0437	&	0.0090	&	0.0079	&	0.0020	&	0.0079	&	0.0079	&	0.0078	\\
			&		&		&		&		&		&		&		&		&		\\
			\\[-1.6ex] \multicolumn{2}{c}{$\delta = 0.02$} &&&&&&&& \\	\cmidrule(lr){1-2}
			$n = 500$	&	MSE	&	1.0000	&	3.1695	&	1.9305	&	1.9422	&	0.5044	&	1.9422	&	1.9291	&	1.9267	\\
			&	|BIAS|	&	0.0061	&	0.0410	&	0.0132	&	0.0125	&	0.0037	&	0.0125	&	0.0124	&	0.0124	\\
			&		&		&		&		&		&		&		&		&		\\
			$n = 1000$	&	MSE	&	1.0000	&	2.2236	&	1.8835	&	1.9013	&	0.5061	&	1.9013	&	1.8925	&	1.8915	\\
			&	|BIAS|	&	0.0025	&	0.0150	&	0.0058	&	0.0061	&	0.0014	&	0.0061	&	0.0063	&	0.0063	\\
			&		&		&		&		&		&		&		&		&		\\
			\\[-1.6ex] \multicolumn{2}{c}{$\delta = 0.05$} &&&&&&&& \\	\cmidrule(lr){1-2}
			$n = 500$	&	MSE	&	1.0000	&	1.4475	&	1.3212	&	1.3292	&	0.6076	&	1.3292	&	1.3243	&	1.3236	\\
			&	|BIAS|	&	0.0068	&	0.0141	&	0.0085	&	0.0085	&	0.0030	&	0.0085	&	0.0085	&	0.0085	\\
			&		&		&		&		&		&		&		&		&		\\
			$n = 1000$	&	MSE	&	1.0000	&	1.3882	&	1.3064	&	1.3104	&	0.6045	&	1.3104	&	1.3070	&	1.3066	\\
			&	|BIAS|	&	0.0021	&	0.0050	&	0.0023	&	0.0024	&	0.0011	&	0.0024	&	0.0027	&	0.0027	\\ \\ \hline		
		\end{tabularx}
		\begin{minipage}{\textwidth}
			\begin{justify} 
				$\delta$ determines the strength of overlap  by setting bound $(\delta,1-\delta)$ for the true instrument propensity scores. $n$ is the sample size. Mean squared errors (MSE) are all relative to the MSE of the instrumental variables estimator (IV). Empty entries refer to the results of a simulation that could not be conducted due to multicollinearity or other convergence problems. Results are based on 20000 Monte Carlo replications. 
			\end{justify}
		\end{minipage}
	\end{table}
	\begin{table} \scriptsize
		\centering \caption{Design B: Relative Mean Squared Errors and Absolute Biases} \label{tab_BAL_MC_DesignB_1}
		\begin{tabularx}{\textwidth}{ll *{8}{Y}}
			\hline \hline \\
			&		&	IV	&	MLE	&	MLE(2)	&	B($X$)	&	B($D$)	&	B($D$,$X$)	&	B($\hat{D}$)	&	B($\hat{D}_m$)	\\ \hline 
			\\[-1.6ex] \multicolumn{2}{c}{$\delta = 0.01$} &&&&&&&& \\	\cmidrule(lr){1-2}																		
			$n = 500$	&	MSE	&	1.0000	&	8279.1000	&	35.9510	&	2.6970	&	2.3924	&	     $\cdots$	&	2.3656	&	2.2798	\\
			&	|BIAS|	&	0.0611	&	0.0701	&	0.0190	&	0.0234	&	0.0231	&	     $\cdots$	&	0.0223	&	0.0218	\\
			&		&		&		&		&		&		&		&		&		\\
			$n = 1000$	&	MSE	&	1.0000	&	20.1634	&	2.5671	&	2.5270	&	2.2154	&	3.2025	&	2.2046	&	2.1188	\\
			&	|BIAS|	&	0.0564	&	0.0380	&	0.0070	&	0.0077	&	0.0089	&	0.0072	&	0.0087	&	0.0089	\\
			&		&		&		&		&		&		&		&		&		\\
			\\[-1.6ex] \multicolumn{2}{c}{$\delta = 0.02$} &&&&&&&& \\	\cmidrule(lr){1-2}																			
			$n = 500$	&	MSE	&	1.0000	&	29.9923	&	1.9067	&	1.8873	&	1.7314	&	   $\cdots$	&	1.7147	&	1.6688	\\
			&	|BIAS|	&	0.0526	&	0.0336	&	0.0139	&	0.0152	&	0.0160	&	   $\cdots$	&	0.0153	&	0.0152	\\
			&		&		&		&		&		&		&		&		&		\\
			$n = 1000$	&	MSE	&	1.0000	&	2.3012	&	1.8002	&	1.8009	&	1.6456	&	2.0001	&	1.6395	&	1.5948	\\
			&	|BIAS|	&	0.0470	&	0.0174	&	0.0054	&	0.0057	&	0.0069	&	0.0052	&	0.0068	&	0.0070	\\
			&		&		&		&		&		&		&		&		&		\\
			\\[-1.6ex] \multicolumn{2}{c}{$\delta = 0.05$} &&&&&&&& \\	\cmidrule(lr){1-2}																		
			$n = 500$	&	MSE	&	1.0000	&	1.4798	&	1.3218	&	1.3281	&	1.2741	&	1.4044	&	1.2679	&	1.2505	\\
			&	|BIAS|	&	0.0369	&	0.0139	&	0.0068	&	0.0072	&	0.0082	&	0.0068	&	0.0079	&	0.0080	\\
			&		&		&		&		&		&		&		&		&		\\
			$n = 1000$	&	MSE	&	1.0000	&	1.3709	&	1.2769	&	1.2783	&	1.2281	&	1.3107	&	1.2242	&	1.2076	\\
			&	|BIAS|	&	0.0336	&	0.0071	&	0.0039	&	0.0040	&	0.0050	&	0.0038	&	0.0050	&	0.0052	\\ \\ \hline
		\end{tabularx}
		\begin{minipage}{\textwidth}
			\begin{justify} 
				$\delta$ determines the strength of overlap  by setting bound $(\delta,1-\delta)$ for the true instrument propensity scores. $n$ is the sample size. Mean squared errors (MSE) are all relative to the MSE of the instrumental variables estimator (IV). Empty entries refer to the results of a simulation that could not be conducted due to multicollinearity or other convergence problems. Results are based on 20000 Monte Carlo replications. 
			\end{justify}
		\end{minipage}
	\end{table}
	\begin{table} \scriptsize
		\centering \caption{Design C: Relative Mean Squared Errors and Absolute Biases} \label{tab_BAL_MC_DesignC_1}
		\begin{tabularx}{\textwidth}{ll *{8}{Y}}
			\hline \hline \\
			&		&	IV	&	MLE	&	MLE(2)	&	B($X$)	&	B($D$)	&	B($D$,$X$)	&	B($\hat{D}$)	&	B($\hat{D}_m$)	\\ \hline 
			\\[-1.6ex] \multicolumn{2}{c}{$\delta = 0.01$} &&&&&&&& \\	\cmidrule(lr){1-2}																		
			$n = 500$	&	MSE	&	1.0000	&	4762.7000	&	10.2364	&	0.7328	&	0.6277	&	      $\cdots$	&	0.6146	&	0.5880	\\
			&	|BIAS|	&	4.7321	&	0.1183	&	0.8877	&	0.0475	&	0.3145	&	      $\cdots$	&	0.4947	&	0.7011	\\
			&		&		&		&		&		&		&		&		&		\\
			$n = 1000$	&	MSE	&	1.0000	&	226.2384	&	1.4826	&	0.4277	&	0.3610	&	0.5703	&	0.3507	&	0.3322	\\
			&	|BIAS|	&	4.6805	&	1.4438	&	0.3254	&	0.0780	&	0.2508	&	0.1955	&	0.3438	&	0.5277	\\
			&		&		&		&		&		&		&		&		&		\\
			\\[-1.6ex] \multicolumn{2}{c}{$\delta = 0.02$} &&&&&&&& \\	\cmidrule(lr){1-2}																		
			$n = 500$	&	MSE	&	1.0000	&	397.0863	&	1.8028	&	0.6438	&	0.5684	&	      $\cdots$	&	0.5441	&	0.5211	\\
			&	|BIAS|	&	3.8752	&	0.2996	&	0.3860	&	0.1093	&	0.1764	&	      $\cdots$	&	0.3229	&	0.4983	\\
			&		&		&		&		&		&		&		&		&		\\
			$n = 1000$	&	MSE	&	1.0000	&	1.3279	&	0.9548	&	0.3717	&	0.3263	&	0.4189	&	0.3120	&	0.2978	\\
			&	|BIAS|	&	3.8584	&	0.2735	&	0.1672	&	0.0674	&	0.1918	&	0.1266	&	0.2688	&	0.4244	\\
			&		&		&		&		&		&		&		&		&		\\
			\\[-1.6ex] \multicolumn{2}{c}{$\delta = 0.05$} &&&&&&&& \\	\cmidrule(lr){1-2}																		
			$n = 500$	&	MSE	&	1.0000	&	1.5293	&	1.1372	&	0.6202	&	0.5804	&	0.6658	&	0.5466	&	0.5281	\\
			&	|BIAS|	&	2.6284	&	0.2239	&	0.0654	&	0.0707	&	0.1129	&	0.0991	&	0.2070	&	0.3331	\\
			&		&		&		&		&		&		&		&		&		\\
			$n = 1000$	&	MSE	&	1.0000	&	0.7532	&	0.6583	&	0.3754	&	0.3511	&	0.3884	&	0.3322	&	0.3219	\\
			&	|BIAS|	&	2.6325	&	0.0698	&	0.0479	&	0.0313	&	0.1442	&	0.0462	&	0.1922	&	0.3057	\\ \\ \hline		
		\end{tabularx}
		\begin{minipage}{\textwidth}
			\begin{justify} 
				$\delta$ determines the strength of overlap  by setting bound $(\delta,1-\delta)$ for the true instrument propensity scores. $n$ is the sample size. Mean squared errors (MSE) are all relative to the MSE of the instrumental variables estimator (IV). Empty entries refer to the results of a simulation that could not be conducted due to multicollinearity or other convergence problems. Results are based on 20000 Monte Carlo replications. 
			\end{justify}
		\end{minipage}
	\end{table}
	
	Overall, the results suggest that the unnormalized MLE approach is outclassed by all other methods due to its large finite sample variance. In general, most balancing approaches outperform the conventional MLE based IPW estimators MLE and MLE(2) by a substantial margin depending on the design. The differences are most pronounced for small sample sizes and small $\delta$. Adding higher-order information is generally helpful to reduce points estimation risk if there are no estimation problems due to a strong correlation between the different balancing variables as for method B($D,X$). While some of the differences between MLE(2) and the balancing approaches are due to a slightly reduced bias, the reduction in variance seems to be the main driver, in particular for designs with strong treatment effect heterogeneity. Unsurprisingly, for all approaches point estimation risk drops with an increase in the sample size.   
	
	For design A, the IV estimator serves as a benchmark as it is correctly specified. The MLE(2) and all balancing approaches except for $B(D)$ are very close in terms of point estimation risk. This is not surprising as under perfect randomization differences in the distributions of observed covariates across instrument levels can only happen by chance. Note that the infeasible $B(D)$ outclasses even the (correctly specified) parametric IV estimator. This is due to the fact that in design A, both the true instrument propensity scores and treatment selection probability are constant and thus the balanced IPW estimator $B(D)$ collapses to the standard Wald estimator. The latter efficiently incorporates the information on independence of the instrument in this homogeneous design and thus yields lower point estimation risk compared to the overparameterized IV.  
	
	For design B, the IV estimator only has a very small bias and thus can serve as a benchmark method. The balancing approaches other than B($D,X$) outperform MLE and, more importantly, MLE(2). The differences are particularly substantial for $n=500$ and $\delta = 0.01$. Here, even the worst balancing approach leads to a reduction in MSE of $92.5\%$ compared to MLE(2). Including higher-order information seems to be beneficial both theoretically (B($D$)) and empirically (B($\hat{D}$) and B($\hat{D}_m$)). In fact, using estimated treatment selection probabilities for balancing seems to be even slightly superior over using true probabilities by a margin of $0$ to $3$ percentage points. Surprisingly, the slightly misspecified model for balancing (B($\hat{D}_m$)) outperforms all other IPW approaches by a margin of at least $1$ to $9$ percentage points. The infeasible B($D,X$) has convergence problems for $n=500$ and $\delta=0.01, 0.02$ due to the strong correlation between $X_i$ and $E[D_i(0)|X_i]$ in the Roy model under normality similar to multicollinearity problems of standard control function approaches in the spirit of \cite{heckman1979sample}. 
	
	For design C, the IV estimator will be severely biased due to the nonlinear potential outcome mean. Here, the differences between balancing and MLE based methods are most pronounced. In fact, all balancing approaches lead to a substantial reduction in MSE compared to IV, MLE and MLE(2). In fact, the consistent MLE(2) needs a much larger sample size to make up for its larger variance compared to the biased IV estimator. All balancing approaches, however, outperform IV by 27 to 69 percentage points in terms of MSE. In general, the gains over IV are more pronounced for larger samples and larger values of $\delta$. Balancing methods also outperform MLE(2) by a factor between $1.7$ ($\delta = 0.05$, $n = 1000$) to $17.4$ ($\delta= 0.01$, $n = 500$) in terms of MSE. There is a clear hierarchy between the different balancing and MLE methods similar but not identical to design B, i.e.~in terms of MSE we have that B($\hat{D}_m$) < B($\hat{D}$) < B(D) < B(X) < MLE(2) < MLE. Again, the slightly misspecified model comes out on top with a margin of at least $1$ to $3$ percentage points. As in design B, the infeasible B($D,X$) has convergence problems for $n=500$ and $\delta=0.01, 0.02$. If it is feasible it ranks as the worst balancing approach but still substantially above MLE(2). It is interesting that in this design the superior performance of the balancing approaches compared to MLE(2) is mainly due to a reduction in variance as they sometimes have a larger bias than MLE(2). This is not a contradiction to the bias Proposition \eqref{prop_BAL_unbiased1}, as in this design treatment effects are highly heterogeneous even for the compliers. Therefore, there is no guarantee for a small finite sample bias while the effects on the variance through pushing the inverse probability weights towards more moderate values as outlined in Section \ref{sec_BAL_CondVarDuality} are still in place.    
	
	The simulation results are evidence that balancing is a superior strategy compared to using likelihood instrument propensity scores for the inverse probability weighting estimator for the LATE in finite samples. The main channel is the reduction in variance, in particular for heterogeneous designs and small sample sizes. This is in line with the results in the literature on balancing weights for selection on observables, see e.g.~\cite{imai2014covariate}.  Moreover, if one is willing to model the treatment selection process, then including higher-order information through estimates of $E[D_i(0)|X_i]$ as balancing constraints seems to be beneficial for estimation of the LATE even under mild misspecification. However, if the information that enters the treatment step is heavily correlated with other regressors used for balancing, then using the generated regressors only is superior to incorporating them jointly in all designs considered. In applied research, the severity can be examined by looking at the Hessian of the empirical (negative) tailored loss function. Alternatively, all information can be used jointly with regularization as suggested in Section \ref{sec_BAL_regularization1}.

	\section{Empirical Application} \label{sec_ApplicationLATE}
	This section contains an illustration of the empirical balancing method to the case of one-sided noncompliance. We re-evaluate of the causal effect of 401(k) retirement plans on private net total financial assets. 401(k) plans are employer-provided, tax-deferred saving plans with partially matched monetary contributions of the employee. The question is whether these plans help to increase net private savings or asset holdings. The evaluation of the effect is complicated by both observed and unobserved individual heterogeneity such as income levels or preferences that might be related to both asset level and the propensity to select the plan. This implies that a simple comparison of savings between participating and non-participating units is likely to yield a biased estimate for the causal effect. We use 401(k) eligibility as as a conditionally independent instrument for participation as the former is only provided through the employer. Identification then rests on the assumption that eligibility is independent of possible confounding unobserved heterogeneity after conditioning on a set of observed characteristics such as income and other financial and socioeconomic background variables. This and similar identification strategies are widely used in the literature \citep{benjamin2003does,abadie2003semiparametric,chernozhukov2004effects,chernozhukov2017double}, see also \cite{poterba1995401} for a critical assessment. 
	
	The data is an excerpt from the	Survey of Income and Program Participation (SIPP) of 1991.
	We apply the same sample restrictions as in \cite{abadie2003semiparametric}. The final data set contains 9275 observations. The outcome is measured as total net financial assets in US dollars. The treatment variable is an indicator for participation in a 401(k) plan. The instrument is a binary indicator for eligibility. Observed confounding variables are age, income, family size, education, and other financial and socioeconomic background variables as in \cite{chernozhukov2017double}. Note that instead of using income measured in dollars we use the natural logarithm of income instead.\footnote{Income has a heavily right-skewed distribution that leads to instability in the nonparametric instrument propensity score estimation step. The log transformation stabilizes the distribution and avoids extreme propensity scores as a result of heavy extrapolation to extremely high income levels.}
	
	We estimate the causal effect of 401(k) participation on net total financial assets using two inverse probability weighting approaches. We also provide standard Wald estimates and replicate the IV result with covariates as in \cite{abadie2003semiparametric} for comparison. For the estimators using inverse probability weights we consider both a simple logistic model and the semiparametric empirical balancing method for estimation of the instrument propensity scores. For both methods we include all possible interactions of the binary regressors. For the empirical balancing method we employ an additive spline basis with degree and number of nodes selected via leave-one-out cross-validation. The loss function used in the cross-validation is the tailored loss function in \eqref{eq_BalancingMAXPROB1}, i.e.~the model is chosen to be optimal in terms of its out-of-sample balance. We also experimented with a non-additive tensor basis but it performed worse in the out-of-sample evaluation. 
	
	\begin{figure} \caption{Estimated Instrument Propensity Score Distributions} 
		\setcounter{figure}{0}
		\begin{subfigure}{.5\textwidth}
			\centering
			\includegraphics[width=\linewidth, trim = 0 100 0 80, clip]{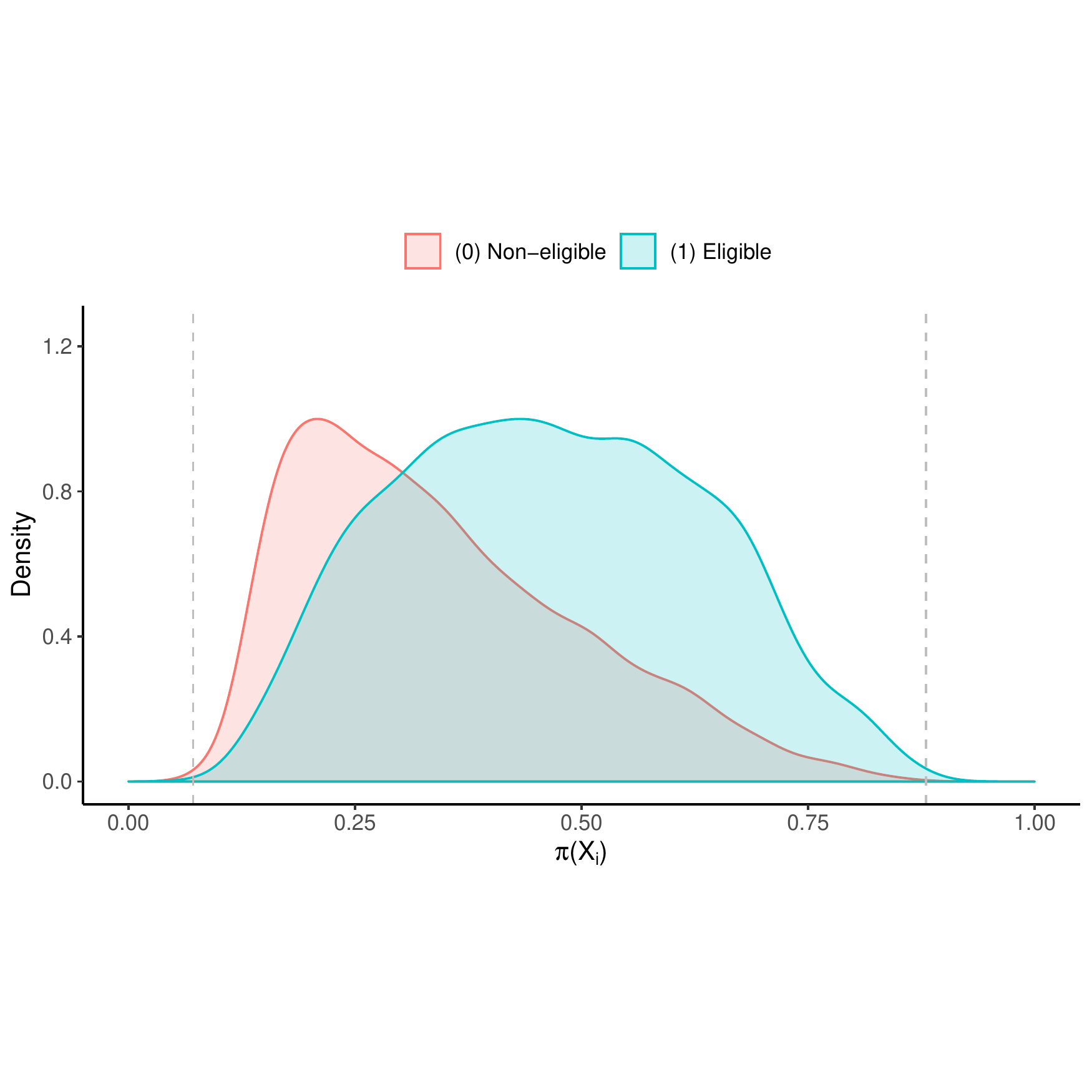}  
			\caption{Logistic Model}
			\label{fig_hist_plog1}
		\end{subfigure}
		\begin{subfigure}{.5\textwidth}
		\centering
		\includegraphics[width=\linewidth, trim = 0 100 0 80, clip]{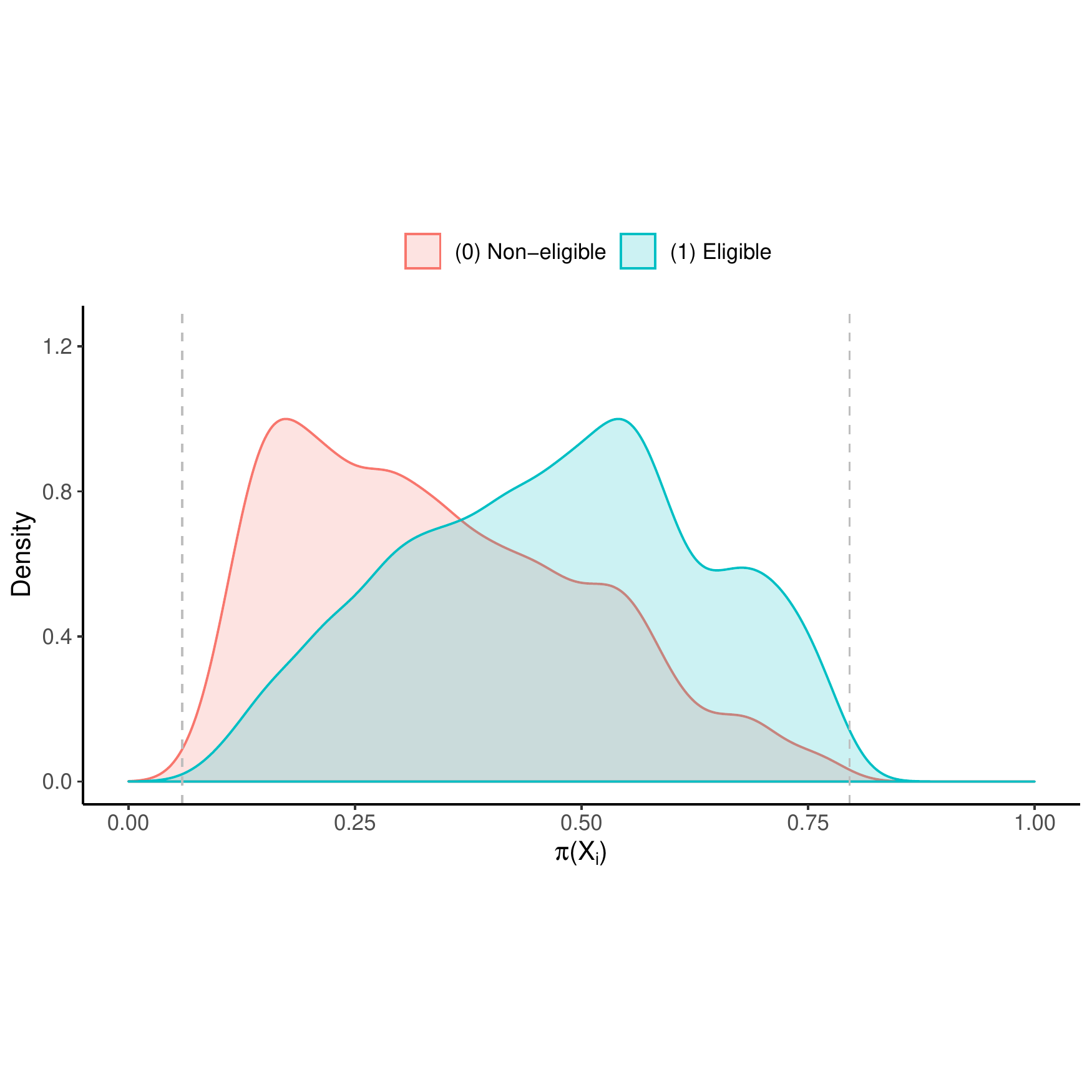}  
			\caption{Empirical Balancing}
			\label{fig_hist_pbal2}
		\end{subfigure}
	{\scriptsize \begin{justify} Conditional density estimates of the instrument propensity scores by eligibility status. The dashed gray lines indicate the minimum and maximum estimated scores for both models. \end{justify} }
	\end{figure}
	
	Figures \ref{fig_hist_plog1} and \ref{fig_hist_pbal2} contain the kernel density estimates  of the instrument propensity scores for eligible and non-eligible units for both the logistic and the empirical balancing scores. Contrary to the more restrictive logistic model, the semiparametric balancing estimator seems to suggest an almost bimodal distribution for the eligible units with more probability mass around probabilities of 0.65 to 0.75, less between 0.25 and 0.5, and more extreme propensities in the left tail. One can see that for both models there is sufficient overlap in the distribution and minimum scores are sufficiently far away from the boundary. Thus we do not expect any irregularity problems regarding the identification of the treatment effect parameters \citep{khan2010irregular,heiler2020inference}. 

	\begin{table} \centering \caption{LATE Estimation Results} \label{tab_LATEresults}
		\begin{tabularx}{\textwidth}{lYYYY} \\[-1ex] \hline \hline 
			      			&	Wald   &   IV 			&      IPW  &    IPW  \\
			     			&     	   & (\textit{Abadie}) &(\textit{Likelihood})  	& (\textit{Balancing}) \\[1ex] \hline  \\[-1ex]
			Estimate 		& 26771.16 &9418.83 		&9558.01 	&12502.77 \\
			Standard Error  & 2023.04  &2152.08 		&1896.57 	&1652.82 \\[1ex] \hline \hline 
		\end{tabularx}
	{\scriptsize \begin{justify} For Wald and IV, standard errors are calculated using standard heteroskedasticity robust estimators. For both IPW methods, standard errors are obtained via  nonparametric bootstrap using 500 replications. All estimates are significant on a 1\% significance level.  \end{justify} }
	\end{table}

	Table \ref{tab_LATEresults} contains the point estimates of the causal effect and the corresponding standard errors for all methods.  The estimates are all significant on a 1\% significance level and broadly consistent with the results previously reported in the literature. The Wald estimate of 26771\$ differs severely from the other estimates as it is based on the overly restrictive identification assumption of unconditionally independent eligibility. The IV estimates using the model by \cite{abadie2003semiparametric} and the parametric IPW estimator yield a very similar result of an around 9500\$ increase in total net financial assets. The more robust semiparametric balancing estimator suggest about a 30\% larger effect. Note that, despite its higher flexibility, empirical balancing has a lower standard error than the more restricted logistic model. This is a consequence of the variance stabilizing effect outlined in Section \ref{sec_BAL_CondVarDuality}. Empirical balancing reduces the presence of extreme propensity scores that can lead to extreme weights which heavily affect the point estimates in finite samples.

	\section{Concluding Remarks} \label{sec_BAL_conclusion1}
	In this paper we develop an estimation method for the local average treatment effect that relies on empirical balancing constraints to improve the internal validity of the causal estimand. It has favorable bias and variance properties compared to conventional approaches, is asymptotically normal, and reaches the semiparametric efficiency bound if a sufficiently flexible model is used. Moreover, it does not rely on the use of outcome or treatment selection information and is easy to implement. Monte Carlo simulations suggest that the theoretical advantages translate well to finite samples. 
	
	As the estimator for the local average treatment effect has the typical ratio form, future work should be done to see whether finite sample bias can be further reduced by imposing alternative balancing constraints that do not have to go through two separate reduced form estimates but instead minimize the bias of the ratio directly. In addition, the empirical performance of different modifications and extensions require further attention. For example, using different instrument propensities for the two reduced form type components that also require different normalization schemes could be beneficial in designs with strong heterogeneities in the potential outcomes.  Moreover, the simple balancing approach that uses higher-order information and basic covariates together used in the Monte Carlo study could be combined with e.g.~$\ell_2$-regularization to overcome multicollinearity problems. An additional contribution would be to derive conditions required for the higher-order approaches to conduct semiparametric inference in the presence of generated regressors from nonparametric models.     
	\newpage
	\begin{appendices}
		\section{}
		\subsection{Proof of Theorem \ref{theorem_BAL_efficient1}} \label{app_BAL_proofTheoremEff1}
		\subsubsection{Asymptotically Linear Representation} 
		We use Lemma 1 and 2 from \cite{hirano2003efficient} (denoted as HIR throughout) for the two components of the LATE to obtain asymptotically linear representations. The adjustments required for Lemma 1 and Lemma 2 of HIR shown below fundamentally rely on the global concavity of the tailored (negative) loss functions, see also the online supplement of \cite{zhao2019covariate} for a sketch of this argument.  Note that the results are not uniform in $K$.
		Lemma 1 and 2 from HIR yields the asymptotically linear representations:
		\begin{align*}
		\sqrt{n}(\hat{\Delta} - \Delta) = \frac{1}{\sqrt{n}}\sum_{i=1}^{n}\delta(Y_i,X_i,Z_i) + o_p(1) \\
		\sqrt{n}(\hat{\Gamma} - \Gamma) = \frac{1}{\sqrt{n}}\sum_{i=1}^{n}\gamma(D_i,X_i,Z_i) + o_p(1) \\
		\end{align*}
		with \begin{align*}
		\delta(Y_i,X_i,Z_i) &= \frac{Z_iY_i}{\pi(X_i)} - \frac{(1-Z_i)Y_i}{\pi(X_i)} - \Delta - (Z_i - \pi(X_i))\bigg(\frac{m_1(X_i)}{\pi(X_i)} + \frac{m_0(X_i)}{1-\pi(X_i)}\bigg) \\
		\gamma(D_i,X_i,Z_i) &= \frac{Z_iD_i}{\pi(X_i)} - \frac{(1-Z_i)D_i}{\pi(X_i)} - \Gamma - (Z_i - \pi(X_i))\bigg(\frac{\mu_1(X_i)}{\pi(X_i)} + \frac{\mu_0(X_i)}{1-\pi(X_i)}\bigg). \\
		\end{align*}
		As Assumption A.4 rules out a zero denominator for the LATE with probability going to one, one obtains \begin{align*}
		\sqrt{n}(\hat{\tau} - \tau) = \sqrt{n}\bigg(\frac{\hat{\Delta}}{\hat{\Gamma}} - \frac{\Delta}{\Gamma}\bigg) 
		= \frac{1}{\Gamma}\sqrt{n}(\hat{\Delta}- \Delta) - \frac{\tau}{\Gamma}\sqrt{n}(\hat{\Gamma}-\Gamma)+ o_p(1)
		\end{align*}
		as for the case of using the standard logistic series estimator, see \cite{donald2014testing}. Asymptotic normality follows from the standard CLT. The asymptotic variance can be derived from the expected squares of the sum of the leading terms of the linearized estimators \citep{hahn1998role,frolich2007nonparametric}. 
		\subsection{Adaptation of Lemma 1 in \cite{hirano2003efficient}}
		Let $L(\phi^K(x)'\theta) = 1/(1+\exp(-\phi^K(x)'\theta))$ denote the logistic instrument score using $K$ basis functions. The negative tailored loss function is given by \begin{align*}
		S_n(\theta) &= \frac{1}{n}\sum_{i=1}^{n}S(Z_i,X_i,\theta) \\
		&= \frac{1}{n}\sum_{i=1}^{n}(2Z_i - 1)\ln\bigg(\frac{L(\phi^K(X_i)'\theta)}{1-L(\phi^K(X_i)'\theta)}\bigg) \\&\quad + (Z_i - L(\phi^K(X_i)'\theta))\bigg(\frac{1}{L(\phi^K(X_i)'\theta)} - \frac{1}{1-L(\phi^K(X_i)'\theta)}\bigg).
		\end{align*} 
		Let the expected loss and pseudo loss for a given $K$ be defined as\begin{align*}
		Q^*(\theta) &= E[S(\pi(X_i),X_i,\theta)] \\
		Q_K(\theta) &= E[S(L(\phi^K(X_i)'\theta_K),X_i,\theta)]
		\end{align*}with maximizers \begin{align*}
		\theta_K^* &= \arg\underset{\theta}{\max}\ Q^*(\theta) \\
		\theta_K &= \arg\underset{\theta}{\max}\ Q_K(\theta).
		\end{align*}
		Now define the compact set with bounded instrument propensity scores \begin{align*}
		\Theta_K = \{\theta \in \mathbb{R}^K: \inf\{\underset{x\in\mathcal{X}}{\inf}L(\phi^K(x)'\theta),\underset{x\in\mathcal{X}}{\inf}(1-L(\phi^K(x)'\theta))\} \geq \eta/2\}
		\end{align*}
		for $\eta = \inf\{\underset{x\in\mathcal{X}}{\inf}\pi(x),\underset{x\in\mathcal{X}}{\inf}(1-\pi(x))\} > 0$ by strong instrument overlap. By the series approximation and the monotonicity of the logistic link it follows as in HIR that \begin{align}
		\underset{x\in\mathcal{X}}{\sup}|\pi(x) - L(\phi^K(x)'\theta_K)| < CK^{-q/r}. \label{app_eq_lemma1_approx1}
		\end{align}Thus for large $K$ we have that $\theta_K \in \Theta_K$. Moreover, note that the difference between negative expected and expected pseudo loss is given by \begin{align*}
		Q^*(\theta) - Q_K(\theta) &= E\bigg[2(\phi(X_i)-L(\phi^K(X_i)'\theta))\ln\bigg(\frac{L(\phi^K(X_i)'\theta)}{1-L(\phi^K(X_i)'\theta)} \\ &\quad  - (\pi(X_i)-L(\phi^K(X_i)'\theta))\bigg(\frac{1}{L(\phi^K(X_i)'\theta)} - \frac{1}{1-L(\phi^K(X_i)'\theta)}\bigg)\bigg)\bigg]
		\end{align*}
		which for $\theta \in \Theta_K$ is bounded by a universal constant depending on $\eta$ from \ref{app_eq_lemma1_approx1} and continuity of the natural logarithm and the inverse. Thus, there exists a constant $C_1$ such that \begin{align*}
		\underset{\theta\in\Theta_K}{\sup}|Q^*(\theta) - Q_K(\theta)| \leq C_1K^{-q/r}.
		\end{align*}  
		Now define a compact set around the maximizer of the pseudo true negative loss \begin{align*}
		\tilde{\Theta}_K =\{\theta\in\mathbb{R}^K: ||\theta - \theta_K|| \leq C_2K^{-q/(2r)} \}.
		\end{align*}
		By following HIR (page 1179), one can show that for $K$ large it must hold that $\tilde{\Theta}_K \subset \Theta_K$. A mean-value expansion of the difference of the expected tailored loss around the optimum yields \begin{align*}
		Q^*(\theta_K) - Q^*(\theta) \geq \frac{\partial Q_K(\theta_K)}{\partial{\theta}'}(\theta-\theta_K) - \frac{1}{2}(\theta-\theta_K)'\frac{\partial^2 Q_K(\tilde{\theta})}{\partial\theta\partial\theta'}(\theta-\theta_K) - 2C_1K^{-q/r} 
		\end{align*}
		with the first expression on the right-hand side being equal to zero by definition. To bound the second term, note that $\tilde{\theta}$ is on the line segment between $\theta$ and $\theta_K$ and thus $\tilde{\theta} \in \tilde{\Theta}_K \subset \Theta_K$. On this set, the tailored loss function is bounded and thus the derivative of its expectation is given by \begin{align*}
		\frac{\partial Q_K(\theta)}{\partial\theta} &= E\bigg[\frac{\partial S(L(\phi^K(X_i)'\theta_K),X_i,\theta)}{\partial\theta}\bigg] \\
		&= E\bigg[\bigg(\frac{L(\phi^K(X_i)'\theta_K)}{L(\phi^K(X_i)'\theta)}  - \frac{1-L(\phi^K(X_i)'\theta_K)}{1-L(\phi^K(X_i)'\theta)}\bigg)\phi^K(X_i)\bigg].
		\end{align*}  
		Regarding the second derivative note again that $\Theta_K$ is compact and $X_i$ compactly supported which implies a bounded first derivative and therefore \begin{align*}
		\frac{\partial^2 Q_K(\theta)}{\partial\theta\partial\theta'} &= -E\bigg[\bigg(\frac{L(\phi^K(X_i)'\theta_K)}{L(\phi^K(X_i)'\theta)}(1-L(\phi^K(X_i)'\theta)) \\&\quad  - \frac{1-L(\phi^K(X_i)'\theta_K)}{1-L(\phi^K(X_i)'\theta)}L(\phi^K(X_i)'\theta)\bigg)\phi^K(X_i)\phi^K(X_i)'\bigg]. 
		\end{align*} 
		Moreover, since the probabilities are nonnegative and below one and $\theta_K \in \Theta_K$ it follows that \begin{align*}
		-\frac{\partial^2 Q_K(\theta)}{\partial\theta\partial\theta'} &\geq E[\inf\{\underset{x\in\mathcal{X}}{\inf}L(\phi^K(x)'\theta),\underset{x\in\mathcal{X}}{\inf}(1-L(\phi^K(x)'\theta))\}\phi^K(X_i)\phi(X_i)'] \\
		&\geq \eta/2 E[\phi(X_i)\phi(X_i)'] \\ &= \eta/2 I_K.
		\end{align*}
		Thus, we have that the minimal eigenvalue of the negative Hessian is also bounded from below \begin{align*}
		\lambda_{min}\bigg(-\frac{\partial^2 Q_K(\theta)}{\partial\theta\partial\theta'}\bigg) \geq \eta/2 \lambda_{min}(I_K) = \eta/2.
		\end{align*}
		Which then implies that for $\theta \neq \theta_K$ \begin{align*}
		Q^*(\theta_K) - Q^*(\theta) > 0. 
		\end{align*}
		As the expected tailored loss is globally concave like the expected binary likelihood with logistic link in HIR, Lemma 1 from HIR then follows equivalently (see HIR, page 1179). 
		
		\subsection{Adaptation of Lemma 2 in \cite{hirano2003efficient}}
		First, we need to show that the derivate of the tailored loss evaluated at the optimum of the expected loss is bounded at the same rate as in HIR, i.e. \begin{align}
		\frac{\partial S_n(\theta_K^*)}{\partial\theta} = O_p(\sqrt{K/n}). \label{app_eq_orderscore1}
		\end{align}
		Note that by the derivations for Lemma 1, we know that $\theta_K^* \in \tilde{\Theta}_K \subset \Theta_K$ and thus by independence
		\begin{align*}
		E\bigg[\bigg|\bigg|&\frac{\partial S_n(\theta_K^*)}{\partial \theta}\bigg|\bigg|^2\bigg] = \frac{1}{n}tr\bigg(E\bigg[\frac{\partial S(\theta_K^*)}{\partial \theta}\frac{\partial S(\theta_K^*)}{\partial \theta'}\bigg]\bigg) \\
		&= \frac{1}{n}tr\bigg(E\bigg[\bigg(\frac{\pi(X_i)}{L(\phi^K(X_i)'\theta_K^*)^2} + \frac{1-\pi(X_i)}{(1-L(\phi^K(X_i)'\theta_K^*))^2}\bigg)\phi^K(X_i)\phi^K(X_i)'\bigg]\bigg) \\
		&\leq \frac{2}{n}tr\bigg(E\bigg[\underset{x\in \mathcal{X}}{\sup}\bigg{\{}\frac{1}{L(\phi^K(x)'\theta_K^*)^2},\frac{1}{(1-L(\phi^K(x)'\theta_K^*))^2}\bigg{\}}\phi^K(X_i)\phi^K(X_i)'\bigg]\bigg) \\
		&\leq \frac{2}{n \eta^2}tr(E[\phi^K(X_i)\phi^K(X_i)']) \\
		&= \frac{2}{\eta^2}\frac{K}{n}.
		\end{align*}
		$\eta$ is a universal constant by strong instrument overlap thus \eqref{app_eq_orderscore1} follows from Markov's inequality. Now we need  to show that with arbitrarily high probability, the likelihood evaluated at $\theta_K^*$ is strictly larger than any other $\theta$ for $||\theta - \theta_K^*|| = C\sqrt{K/n}$ as in HIR. The proof simplifies due to the different structure of the Hessian compared to the likelihood case in HIR. A mean value expansion of the tailored loss yields \begin{align*}
		S_n(\theta) - S_n(\theta^*_K) = \frac{\partial S_n(\theta_K^*)}{\partial\theta'}(\theta - \theta_K) + \frac{1}{2}(\theta - \theta_K)'\frac{\partial^2 S_n(\tilde{\theta})}{\partial\theta\partial\theta'}(\theta - \theta_K)
		\end{align*}
		with $\tilde{\theta}$ being on the line segment between $\theta$ and $\theta_K$. The expected negative Hessian at the intermediate for a fixed $K$ is given by \begin{align*}
		-E\bigg[\frac{\partial^2 S_n(\tilde{\theta})}{\partial\theta\partial\theta'}\bigg] &= E\bigg[\bigg(\frac{\pi(X_i)}{L(\phi^K(X_i)'\tilde{\theta})}(1-L(\phi^K(X_i)'\tilde{\theta})) \\&\quad  + \frac{1-\pi(X_i)}{1-L(\phi^K(X_i)'\tilde{\theta})}L(\phi^K(X_i)'\tilde{\theta})\bigg)\phi^K(X_i)\phi^K(X_i)'\bigg] \\
		&\geq E[(\pi(X_i)(1-L(\phi^K(X_i)'\tilde{\theta}))\\ &\quad  + (1-\pi(X_i))L(\phi^K(X_i)'\tilde{\theta}))\phi^K(X_i)\phi^K(X_i)'] \\
		&\geq \inf\{\underset{x\in\mathcal{X}}{\inf}\pi(x),\underset{x\in\mathcal{X}}{\inf}(1-\pi(x))\} E[\phi^K(X_i)\phi^K(X_i)] \\
		&\geq \frac{\eta}{2}I_K
		\end{align*} 
		since $\tilde{\theta}$ is on the line-segment between $\theta$ and $\theta_K$ and thus $\tilde{\theta} \in \tilde{\Theta}_K$. Since the right-hand side is also bounded from above by the equivalent argument as for the Hessian at $\theta_K^*$, it implies that for a fixed $K$\begin{align*}
		-\frac{\partial^2 S_n(\tilde{\theta})}{\partial\theta\partial\theta'} \geq  \frac{\eta}{2}I_K + o_p(1).
		\end{align*}
		Thus, we can find a sample size $n_0$ such that with probability of at least $1-\varepsilon/2$ for $n\geq n_0$ it holds that  ${ \lambda_{min}(\partial^2S_n(\tilde{\theta})}/{\partial\theta\partial\theta'}) \geq  {\eta}/{2}$. Now, using \eqref{app_eq_orderscore1} choose $n \geq n_0$ large and a constant $C$ such that \begin{align*}
		P\bigg(\bigg|\bigg|\frac{\partial S_n(\theta_K^*)}{\partial\theta}\bigg|\bigg| < \frac{\eta}{4}C\sqrt{\frac{K}{n}}\bigg) \geq 1-\varepsilon/2.
		\end{align*}
		Thus, for $K$ and $n$ large we have with probability of at least $1-\varepsilon$
		\begin{align*}
		S_n(\theta) - S_n(\theta_K^*) &\leq \bigg|\bigg|\frac{\partial S_n(\theta_K^*)}{\partial\theta}\bigg|\bigg| ||\theta - \theta_K|| - \frac{\eta}{4}||\theta - \theta_K^*||^2 \\
		&\leq\bigg(\bigg|\bigg|\frac{\partial S_n(\theta_K^*)}{\partial\theta}\bigg|\bigg| - \frac{\eta}{4}C\sqrt{\frac{K}{n}}\bigg)||\theta-\theta_K^*|| \\
		&< 0.
		\end{align*}
		The remaining steps for Lemma 2 follow from continuity and global concavity of the tailored loss function as in HIR, page 1181.
		
		\subsection{Proof of Proposition \ref{prop_BAL_unbiased1}} \label{app_BAL_proofBIASratio}
		We first derive the individual expectations of numerator and denominator using the tailored loss scores. The proposition then follows from the law of iterated expectations. Let $\hat{\pi}(X_i)$ for $i=1,\dots,n$ denote the instrument propensity scores obtained from the tailored loss. We make use of the conditional independence, in particular of the fact that conditional on the full set of $X_1,\dots,X_n$ and $Z_1,\dots,Z_n$ the balanced scores $\hat{\pi}(X_i)$ are deterministic. The potential outcomes and potential treatment levels $D_i(z)$ and $Y_i(d)$ however do not depend on $X_j$ and $Z_j$ for $j\neq i$ by Assumption A.6 and are jointly independent of $Z_i$ conditional on $X_i$ by Assumption A.1, which implies for example that  \begin{align*}
		E\bigg[\frac{Z_i}{\hat{\pi}(X_i)}D_i(1)\bigg|X_1,\dots,X_n,Z_1,\dots,Z_n\bigg] &= E\bigg[\frac{Z_i}{\hat{\pi(X_i)}}E[D_i(1)|X_1,\dots,X_n,Z_1,\dots,Z_n]\bigg] \\ &= E\bigg[\frac{Z_i}{\hat{\pi(X_i)}}E[D_i(1)|X_i]\bigg]. 
		\end{align*}Similarly, in the following derivations, all iterated expectations reduce to conditioning on $X_i$ only. For the remainder of the proof let $\mathcal{S}_{\phi}$ be the linear span of $\phi_1(X_i),\dots,\phi_r(X_i)$ (which can include a constant). 
		\subsubsection{Expectation of the Numerator}
		For the denominator of the balanced LATE estimator we have that {\scriptsize\begin{align*}
			\hat{\Gamma} &= \frac{1}{n}\sum_{i=1}^{n}\frac{Z_i}{\hat{\pi}(X_i)}D_i - \frac{1}{n}\sum_{i=1}^{n}\frac{(1-Z_i)}{1-\hat{\pi}(X_i)}D_i \\
			&=  \frac{1}{n}\sum_{i=1}^{n}\frac{Z_i}{\hat{\pi}(X_i)}(D_i(1)-E[D_i(1)|X_i]) +  \frac{1}{n}\sum_{i=1}^{n}\frac{Z_i}{\hat{\pi}(X_i)}E[D_i(1)|X_i] \\ 
			&\quad - \frac{1}{n}\sum_{i=1}^{n}\frac{1-Z_i}{1-\hat{\pi}(X_i)}(D_i(0)-E[D_i(0)|X_i]) +  \frac{1}{n}\sum_{i=1}^{n}\frac{1-Z_i}{1-\hat{\pi}(X_i)}E[D_i(0)|X_i] \\
			&= \frac{1}{n}\sum_{i=1}^{n}\frac{Z_i}{\hat{\pi}(X_i)}(D_i(1)-E[D_i(1)|X_i]) - \frac{1}{n}\sum_{i=1}^{n}\frac{1-Z_i}{1-\hat{\pi}(X_i)}(D_i(0)-E[D_i(0)|X_i]) \\
			&\quad + \frac{1}{n}\sum_{i=1}^{n}\frac{Z_i}{\hat{\pi}(X_i)}(E[D_i(1)|X_i]-E[D_i(0)|X_i]) +  \frac{1}{n}\sum_{i=1}^{n}\frac{Z_i-\hat{\pi}(X_i)}{\hat{\pi}(X_i)(1-\hat{\pi}(X_i))}E[D_i(0)|X_i] 
			\end{align*}}
		or equivalently 
		{\scriptsize\begin{align*}
			\hat{\Gamma} &= \frac{1}{n}\sum_{i=1}^{n}\frac{Z_i}{\hat{\pi}(X_i)}(D_i(1)-E[D_i(1)|X_i]) - \frac{1}{n}\sum_{i=1}^{n}\frac{1-Z_i}{1-\hat{\pi}(X_i)}(D_i(0)-E[D_i(0)|X_i]) \\
			&\quad + \frac{1}{n}\sum_{i=1}^{n}\frac{1-Z_i}{1-\hat{\pi}(X_i)}(E[D_i(1)|X_i]-E[D_i(0)|X_i]) +  \frac{1}{n}\sum_{i=1}^{n}\frac{Z_i- \hat{\pi}(X_i)}{\hat{\pi}(X_i)(1-\hat{\pi}(X_i))}E[D_i(1)|X_i]. 
			\end{align*}}
		Thus, if $E[D_i(0)|X_i] = E[D_i|X_i,Z_i=0] \in \mathcal{S}_{\phi}$ then {\scriptsize\begin{align*}
			E[\hat{\Gamma}] = E\bigg[ \frac{1}{n}\sum_{i=1}^{n}\frac{Z_i}{\hat{\pi}(X_i)}(E[D_i(1)|X_i]-E[D_i(0)|X_i])\bigg]
			\end{align*}}
		and equivalently if $E[D_i(1)|X_i] = E[D_i|X_i,Z_i=1] \in \mathcal{S}_{\phi}$ then {\scriptsize\begin{align*}
			E[\hat{\Gamma}] = E\bigg[ \frac{1}{n}\sum_{i=1}^{n}\frac{1-Z_i}{1-\hat{\pi}(X_i)}(E[D_i(1)|X_i]-E[D_i(0)|X_i])\bigg].
			\end{align*}}
		\subsubsection{Expectation of the Denominator}
		{\scriptsize\begin{align*}
			\hat{\Delta} &= \frac{1}{n}\sum_{i=1}^{n}\frac{Z_i}{\hat{\pi}(X_i)}Y_i - \frac{1}{n}\sum_{i=1}^{n}\frac{(1-Z_i)}{1-\hat{\pi}(X_i)}Y_i \\ 
			&= \frac{1}{n}\sum_{i=1}^{n}\frac{Z_i}{\hat{\pi}(X_i)}(Y_i(0) + D_i(1)(Y_i(1)-Y_i(0)) -  \frac{1}{n}\sum_{i=1}^{n}\frac{(1-Z_i)}{1-\hat{\pi}(X_i)}(Y_i(0) + D_i(0)(Y_i(1)-Y_i(0)) \\
			&=\frac{1}{n}\sum_{i=1}^{n}\frac{Z_i-\hat{\pi}(X_i)}{\hat{\pi}(X_i)(1-\hat{\pi}(X_i))}Y_i(0) + \frac{1}{n}\sum_{i=1}^{n}\frac{Z_i}{\hat{\pi}(X_i)}(D_i(1)-D_i(0))(Y_i(1)-Y_i(0)) \\
			&\quad + \frac{1}{n}\sum_{i=1}^{n}\frac{Z_i-\hat{\pi}(X_i)}{\hat{\pi}(X_i)(1-\hat{\pi}(X_i))}D_i(0)(Y_i(1)-Y_i(0)) \\
			&= \frac{1}{n}\sum_{i=1}^{n}\frac{Z_i-\hat{\pi}(X_i)}{\hat{\pi}(X_i)(1-\hat{\pi}(X_i))}E[Y_i(0)|X_i] + \frac{1}{n}\sum_{i=1}^{n}\frac{Z_i-\hat{\pi}(X_i)}{\hat{\pi}(X_i)(1-\hat{\pi}(X_i))}(Y_i(0) - E[Y_i(0)|X_i]) \\
			&\quad + \frac{1}{n}\sum_{i=1}^{n}\frac{Z_i-\hat{\pi}(X_i)}{\hat{\pi}(X_i)(1-\hat{\pi}(X_i))}E[D_i(0)(Y_i(1)-Y_i(0))|X_i] \\ &\quad +\frac{1}{n}\sum_{i=1}^{n}\frac{Z_i-\hat{\pi}(X_i)}{\hat{\pi}(X_i)(1-\hat{\pi}(X_i))}(D_i(0)(Y_i(1)-Y_i(0)) - E[D_i(0)(Y_i(1)-Y_i(0))|X_i]) \\
			&\quad + \frac{1}{n}\sum_{i=1}^{n}\frac{Z_i}{\hat{\pi}(X_i)}(D_i(1)-D_i(0))(Y_i(1)-Y_i(0))
			\end{align*}}
		or equivalently {\scriptsize\begin{align*}
			\hat{\Delta}&= \frac{1}{n}\sum_{i=1}^{n}\frac{Z_i-\hat{\pi}(X_i)}{\hat{\pi}(X_i)(1-\hat{\pi}(X_i))}E[Y_i(0)|X_i] + \frac{1}{n}\sum_{i=1}^{n}\frac{Z_i-\hat{\pi}(X_i)}{\hat{\pi}(X_i)(1-\hat{\pi}(X_i))}(Y_i(0) - E[Y_i(0)|X_i]) \\
			&\quad + \frac{1}{n}\sum_{i=1}^{n}\frac{Z_i-\hat{\pi}(X_i)}{\hat{\pi}(X_i)(1-\hat{\pi}(X_i))}E[D_i(1)(Y_i(1)-Y_i(0))|X_i] \\ &\quad +\frac{1}{n}\sum_{i=1}^{n}\frac{Z_i-\hat{\pi}(X_i)}{\hat{\pi}(X_i)(1-\hat{\pi}(X_i))}(D_i(1)(Y_i(1)-Y_i(0)) - E[D_i(1)(Y_i(1)-Y_i(0))|X_i]) \\
			&\quad + \frac{1}{n}\sum_{i=1}^{n}\frac{1-Z_i}{1-\hat{\pi}(X_i)}(D_i(1)-D_i(0))(Y_i(1)-Y_i(0)).
			\end{align*}}
		Thus, if $E[Y_i(0)|X_i]$, $E[D_i(0)(Y_i(1)-Y_i(0))|X_i] \in \mathcal{S}_{\phi}$ then {\scriptsize\begin{align*}
			E[\hat{\Delta}] &= E\bigg[\frac{1}{n}\sum_{i=1}^{n}\frac{Z_i}{\hat{\pi}(X_i)}(D_i(1)-D_i(0))(Y_i(1)-Y_i(0))\bigg] \\
			&= E\bigg[\frac{1}{n}\sum_{i=1}^{n}\frac{Z_i}{\hat{\pi}(X_i)}E[(D_i(1)-D_i(0))(Y_i(1)-Y_i(0))|X_i]\bigg]
			\end{align*}}
		and equivalently if $E[Y_i(0)|X_i]$, $E[D_i(1)(Y_i(1)-Y_i(0))|X_i] \in \mathcal{S}_{\phi}$ {\scriptsize\begin{align*}
			E[\hat{\Delta}] &= E\bigg[\frac{1}{n}\sum_{i=1}^{n}\frac{1-Z_i}{1-\hat{\pi}(X_i)}(D_i(1)-D_i(0))(Y_i(1)-Y_i(0))\bigg] \\
			&= E\bigg[\frac{1}{n}\sum_{i=1}^{n}\frac{1-Z_i}{1-\hat{\pi}(X_i)}E[(D_i(1)-D_i(0))(Y_i(1)-Y_i(0))|X_i]\bigg].
			\end{align*}}
		\subsubsection{Ratio of Expectations and Homogeneity}
		Thus, we have that if $E[D_i(0)|X_i]$, $E[Y_i(0)|X_i]$, $E[D_i(0)(Y_i(1)-Y_i(0))|X_i] \in \mathcal{S}_{\phi}$ {\scriptsize\begin{align*}
			\frac{E[\hat{\Delta}]}{E[\hat{\Gamma}]} = \frac{E\bigg[\frac{1}{n}\sum_{i=1}^{n}\frac{Z_i}{\hat{\pi}(X_i)}E[(D_i(1)-D_i(0))(Y_i(1)-Y_i(0))|X_i]\bigg]}{E\bigg[ \frac{1}{n}\sum_{i=1}^{n}\frac{Z_i}{\hat{\pi}(X_i)}(E[D_i(1)|X_i]-E[D_i(0)|X_i])\bigg]}
			\end{align*}}
		or equivalently if $E[D_i(1)|X_i]$, $E[Y_i(0)|X_i]$, $E[D_i(1)(Y_i(1)-Y_i(0))|X_i] \in \mathcal{S}_{\phi}$ {\scriptsize\begin{align*}
			\frac{E[\hat{\Delta}]}{E[\hat{\Gamma}]} = \frac{E\bigg[\frac{1}{n}\sum_{i=1}^{n}\frac{1-Z_i}{1-\hat{\pi}(X_i)}E[(D_i(1)-D_i(0))(Y_i(1)-Y_i(0))|X_i]\bigg]}{E\bigg[ \frac{1}{n}\sum_{i=1}^{n}\frac{1-Z_i}{1-\hat{\pi}(X_i)}(E[D_i(1)|X_i]-E[D_i(0)|X_i])\bigg]}.
			\end{align*}}
		Thus, if the conditional LATE is constant, i.e. if $E[(Y_i(1)-Y_i(0))|X_i,D_i(1)>D_i(0)] = \tau$, then since $P(D_i(1)>D_i(0)|X_i) = E[D_i(1)-D_i(0)|X_i]$, it follows from monotonicity that
		{\scriptsize\begin{align*}
			\frac{E[\hat{\Delta}]}{E[\hat{\Gamma}]} &= \frac{E\bigg[\frac{1}{n}\sum_{i=1}^{n}\frac{Z_i}{\hat{\pi}(X_i)}E[(Y_i(1)-Y_i(0))|X_i,D_i(1)>D_i(0)]P(D_i(1)>D_i(0))|X_i] \bigg]}{E\bigg[ \frac{1}{n}\sum_{i=1}^{n}\frac{Z_i}{\hat{\pi}(X_i)}(E[D_i(1)|X_i]-E[D_i(0)|X_i])\bigg]} \\
			&= \tau 
			\end{align*}}
		and equivalently for the other normalization. Thus the balanced LATE estimator is a ratio of two unbiased estimators.  
		
		\subsection{Asymptotic Variance Estimation} \label{sec_BAL_AsymptoticVar}
		The following is taken from \cite{donald2014testing} and adapted to the notation in this paper. It follows from the derivations in Appendix \ref{app_BAL_proofTheoremEff1} that the LATE estimator can be written in an asymptotically linear representation, i.e. 
		\begin{align*}
		\sqrt{n}(\hat{\tau}_{LATE} - \tau_{LATE}) = \frac{1}{\sqrt{n}}\sum_{i=1}^{n}\psi(Y_i,D_i,Z_i,X_i) + o_p(1)
		\end{align*}
		with \begin{align*}
		\psi(Y_i,D_i,Z_i,X_i) &= \frac{1}{\Gamma}\bigg[\frac{Z_i(Y_i - m_1(X_i) - \tau_{LATE}(D_i-\mu_1(X_i))}{\pi(X_i)} \\
		&\quad - \frac{(1-Z_i)(Y_i - m_0(X_i) - \tau_{LATE}(D_i - \mu_0(X_i)))}{1-\pi(X_i)} \\
		&\quad + m_1(X_i) - m_0(X_i) - \tau_{LATE}(\mu_1(X_i) - \mu_0(X_i))\bigg].
		\end{align*}
		The estimator for the asymptotic variance is then given by
		\begin{align*}
		\hat{V} = \frac{1}{n}\sum_{i=1}^{n}\hat{\psi}(Y_i,D_i,Z_i,X_i)^2
		\end{align*}
		with $\hat{\psi}(\cdot)$ corresponding to $\psi(\cdot)$ with all population moments replaced by sample estimates, i.e. \begin{align*}
		\hat{m}_1(X_i) &= \bigg(\sum_{i=1}^{n}\frac{Y_iZ_i}{\hat{\pi}(X_i)}\phi^K(X_i)\bigg)'\bigg(\sum_{i=1}^{n}\phi^K(X_i)\phi^K(X_i)'\bigg)^{-1}\phi^K(X_i) \\
		\hat{m}_0(X_i) &= \bigg(\sum_{i=1}^{n}\frac{Y_i(1-Z_i)}{1-\hat{\pi}(X_i)}\phi^K(X_i)\bigg)'\bigg(\sum_{i=1}^{n}\phi^K(X_i)\phi^K(X_i)'\bigg)^{-1}\phi^K(X_i) \\
		\hat{\mu}_1(X_i) &= \bigg(\sum_{i=1}^{n}\frac{D_iZ_i}{\hat{\pi}(X_i)}\phi^K(X_i)\bigg)'\bigg(\sum_{i=1}^{n}\phi^K(X_i)\phi^K(X_i)'\bigg)^{-1}\phi^K(X_i) \\
		\hat{\mu}_0(X_i) &= \bigg(\sum_{i=1}^{n}\frac{D_i(1-Z_i)}{1-\hat{\pi}(X_i)}\phi^K(X_i)\bigg)'\bigg(\sum_{i=1}^{n}\phi^K(X_i)\phi^K(X_i)'\bigg)^{-1}\phi^K(X_i). 
		\end{align*}
		
	\end{appendices}

\newpage
	\addcontentsline{toc}{section}{References}	
	\bibliography{Balance1}

\begin{thebibliography}{}

\bibitem[\protect\astroncite{Abadie}{2003}]{abadie2003semiparametric}
Abadie, A. (2003).
\newblock {S}emiparametric {I}nstrumental {V}ariable {E}stimation of
  {T}reatment {R}esponse {M}odels.
\newblock {\em {J}ournal of {E}conometrics}, 113(2):231--263.

\bibitem[\protect\astroncite{Angrist}{1990}]{angrist1990lifetime}
Angrist, J.~D. (1990).
\newblock Lifetime {E}arnings and the {V}ietnam {E}ra {D}raft {L}ottery:
  {E}vidence from {S}ocial {S}ecurity.
\newblock {\em {T}he {A}merican {E}conomic {R}eview}, 80(3):313--336.

\bibitem[\protect\astroncite{Angrist and Imbens}{1995}]{angrist1995two}
Angrist, J.~D. and Imbens, G.~W. (1995).
\newblock Two-stage {L}east {S}quares {E}stimation of {A}verage {C}ausal
  {E}ffects in {M}odels with {V}ariable {T}reatment {I}ntensity.
\newblock {\em Journal of the American Statistical Association},
  90(430):431--442.

\bibitem[\protect\astroncite{Angrist et~al.}{1996}]{angrist1996identification}
Angrist, J.~D., Imbens, G.~W., and Rubin, D.~B. (1996).
\newblock {I}dentification of {C}ausal {E}ffects {U}sing {I}nstrumental
  {V}ariables.
\newblock {\em {J}ournal of the {A}merican {S}tatistical {A}ssociation},
  91(434):444--455.

\bibitem[\protect\astroncite{Angrist and Pischke}{2009}]{angrist2009mostly}
Angrist, J.~D. and Pischke, J.-S. (2009).
\newblock {\em Mostly {H}armless {E}conometrics: {A}n {E}mpiricist's
  {C}ompanion}.
\newblock Princeton University Press.

\bibitem[\protect\astroncite{Athey et~al.}{2018}]{athey2018approximate}
Athey, S., Imbens, G.~W., and Wager, S. (2018).
\newblock Approximate {R}esidual {B}alancing: {D}ebiased {I}nference of
  {A}verage {T}reatment {E}ffects in {H}igh {D}imensions.
\newblock {\em Journal of the Royal Statistical Society: Series B (Statistical
  Methodology)}, 80(4):597--623.

\bibitem[\protect\astroncite{Benjamin}{2003}]{benjamin2003does}
Benjamin, D.~J. (2003).
\newblock Does 401 (k) eligibility increase saving? evidence from propensity
  score subclassification.
\newblock {\em Journal of Public Economics}, 87(5-6):1259--1290.

\bibitem[\protect\astroncite{Bloom}{1984}]{bloom1984accounting}
Bloom, H.~S. (1984).
\newblock Accounting for {N}o-shows in {E}xperimental {E}valuation {D}esigns.
\newblock {\em Evaluation {R}eview}, 8(2):225--246.

\bibitem[\protect\astroncite{Busso et~al.}{2014}]{busso2014new}
Busso, M., DiNardo, J., and McCrary, J. (2014).
\newblock New {E}vidence on the {F}inite {S}ample {P}roperties of {P}ropensity
  {S}core {R}eweighting and {M}atching {E}stimators.
\newblock {\em Review of Economics and Statistics}, 96(5):885--897.

\bibitem[\protect\astroncite{Card}{1995}]{card1995using}
Card, D. (1995).
\newblock Using {G}eographic {V}ariation in {C}ollege {P}roximity to {E}stimate
  the {R}eturn to {S}chooling.
\newblock In {\em Aspects of Labour Market Behaviour: Essays in Honour of John
  Vanderkamp}, pages 201--222. University of Toronto Press.

\bibitem[\protect\astroncite{Card}{2001}]{card2001estimating}
Card, D. (2001).
\newblock Estimating the {R}eturn to {S}chooling: {P}rogress on some
  {P}ersistent {E}conometric {P}roblems.
\newblock {\em Econometrica}, 69(5):1127--1160.

\bibitem[\protect\astroncite{Cawley et~al.}{2013}]{cawley2013impact}
Cawley, J., Frisvold, D., and Meyerhoefer, C. (2013).
\newblock The {I}mpact of {P}hysical {E}ducation on {O}besity among
  {E}lementary {S}chool {C}hildren.
\newblock {\em Journal of Health Economics}, 32(4):743--755.

\bibitem[\protect\astroncite{Cawley et~al.}{2007}]{cawley2007impact}
Cawley, J., Meyerhoefer, C., and Newhouse, D. (2007).
\newblock The {I}mpact of {S}tate {P}hysical {E}ducation {R}equirements on
  {Y}outh {P}hysical {A}ctivity and {O}verweight.
\newblock {\em Health Economics}, 16(12):1287--1301.

\bibitem[\protect\astroncite{Chan et~al.}{2016}]{chan2016globally}
Chan, K. C.~G., Yam, S. C.~P., and Zhang, Z. (2016).
\newblock Globally {E}fficient {N}on-{P}arametric {I}nference of {A}verage
  {T}reatment {E}ffects by {E}mpirical {B}alancing {C}alibration {W}eighting.
\newblock {\em Journal of the Royal Statistical Society: Series B (Statistical
  Methodology)}, 78(3):673--700.

\bibitem[\protect\astroncite{Chernozhukov
  et~al.}{2017}]{chernozhukov2017double}
Chernozhukov, V., Chetverikov, D., Demirer, M., Duflo, E., Hansen, C., Newey,
  W., and Robins, J. (2017).
\newblock Double/debiased {M}achine {L}earning for {T}reatment and {S}tructural
  {P}arameters.
\newblock {\em The Econometrics Journal}, 21(1):1--68.

\bibitem[\protect\astroncite{Chernozhukov and
  Hansen}{2004}]{chernozhukov2004effects}
Chernozhukov, V. and Hansen, C. (2004).
\newblock The {E}ffects of 401(k) {P}articipation on the {W}ealth
  {D}istribution: {A}n {I}nstrumental {Q}uantile {R}egression {A}nalysis.
\newblock {\em Review of Economics and Statistics}, 86(3):735--751.

\bibitem[\protect\astroncite{Donald et~al.}{2014a}]{donald2014inverse}
Donald, S.~G., Hsu, Y.-C., and Lieli, R.~P. (2014a).
\newblock Inverse {P}robability {W}eighted {E}stimation of {L}ocal {A}verage
  {T}reatment {E}ffects: {A} {H}igher {O}rder {MSE} {E}xpansion.
\newblock {\em Statistics \& Probability Letters}, 95:132--138.

\bibitem[\protect\astroncite{Donald et~al.}{2014b}]{donald2014testing}
Donald, S.~G., Hsu, Y.-C., and Lieli, R.~P. (2014b).
\newblock Testing the {U}nconfoundedness {A}ssumption via {I}nverse
  {P}robability \allowbreak {W}eighted {E}stimators of {(L)ATT}.
\newblock {\em Journal of Business \& Economic Statistics}, 32(3):395--415.

\bibitem[\protect\astroncite{Fr{\"o}lich}{2007}]{frolich2007nonparametric}
Fr{\"o}lich, M. (2007).
\newblock {N}onparametric {IV} {E}stimation of {L}ocal {A}verage {T}reatment
  {E}ffects with {C}ovariates.
\newblock {\em {J}ournal of {E}conometrics}, 139(1):35--75.

\bibitem[\protect\astroncite{Fr{\"o}lich and
  Melly}{2013}]{frolich2013identification}
Fr{\"o}lich, M. and Melly, B. (2013).
\newblock Identification of {T}reatment {E}ffects on the {T}reated with
  {O}ne-sided {N}on-compliance.
\newblock {\em Econometric {R}eviews}, 32(3):384--414.

\bibitem[\protect\astroncite{Graham et~al.}{2012}]{graham2012inverse}
Graham, B.~S., de~Xavier~Pinto, C.~C., and Egel, D. (2012).
\newblock Inverse {P}robability {T}ilting for {M}oment {C}ondition {M}odels
  with {M}issing {D}ata.
\newblock {\em The Review of Economic Studies}, 79(3):1053--1079.

\bibitem[\protect\astroncite{Hahn}{1998}]{hahn1998role}
Hahn, J. (1998).
\newblock {O}n the {R}ole of the {P}ropensity {S}core in {E}fficient
  {S}emiparametric {E}stimation of {A}verage {T}reatment {E}ffects.
\newblock {\em {E}conometrica}, 66(2):315--331.

\bibitem[\protect\astroncite{Hainmueller}{2012}]{hainmueller2012entropy}
Hainmueller, J. (2012).
\newblock Entropy {B}alancing for {C}ausal {E}ffects: {A} {M}ultivariate
  {R}eweighting {M}ethod to {P}roduce {B}alanced {S}amples in {O}bservational
  {S}tudies.
\newblock {\em Political Analysis}, 20(1):25--46.

\bibitem[\protect\astroncite{Heckman}{1979}]{heckman1979sample}
Heckman, J.~J. (1979).
\newblock {S}ample {S}election {B}ias as a {S}pecification {E}rror.
\newblock {\em Econometrica}, 47(1):153--161.

\bibitem[\protect\astroncite{Heckman and
  Vytlacil}{2005}]{heckman2005structural}
Heckman, J.~J. and Vytlacil, E.~J. (2005).
\newblock Structural {E}quations, {T}reatment {E}ffects, and {E}conometric
  {P}olicy {E}valuation.
\newblock {\em Econometrica}, 73(3):669--738.

\bibitem[\protect\astroncite{Heiler and Kazak}{2020}]{heiler2020inference}
Heiler, P. and Kazak, E. (forthcoming, 2020).
\newblock Valid {I}nference for {T}reatment {E}ffect {P}arameters under
  {I}rregular {I}dentification and {M}any {E}xtreme {P}ropensity {S}cores.
\newblock {\em Journal of Econometrics}.

\bibitem[\protect\astroncite{Heiler and Mareckova}{2018}]{heiler2018shrinkage}
Heiler, P. and Mareckova, J. (2018).
\newblock Shrinkage for {C}ategorical {R}egressors.
\newblock {\em Working {P}aper}.

\bibitem[\protect\astroncite{Hirano et~al.}{2003}]{hirano2003efficient}
Hirano, K., Imbens, G.~W., and Ridder, G. (2003).
\newblock {E}fficient {E}stimation of {A}verage {T}reatment {E}ffects {U}sing
  the {E}stimated {P}ropensity {S}core.
\newblock {\em {E}conometrica}, 71(4):1161--1189.

\bibitem[\protect\astroncite{Holland}{1986}]{holland1986statistics}
Holland, P.~W. (1986).
\newblock {S}tatistics and {C}ausal {I}nference.
\newblock {\em {J}ournal of the {A}merican {S}tatistical {A}ssociation},
  81(396):945--960.

\bibitem[\protect\astroncite{Hong and Nekipelov}{2010}]{hong2010semiparametric}
Hong, H. and Nekipelov, D. (2010).
\newblock Semiparametric {E}fficiency in {N}onlinear {LATE} {M}odels.
\newblock {\em Quantitative Economics}, 1(2):279--304.

\bibitem[\protect\astroncite{Imai and Ratkovic}{2014}]{imai2014covariate}
Imai, K. and Ratkovic, M. (2014).
\newblock Covariate {B}alancing {P}ropensity {S}core.
\newblock {\em Journal of the Royal Statistical Society: Series B (Statistical
  Methodology)}, 76(1):243--263.

\bibitem[\protect\astroncite{Imbens and
  Angrist}{1994}]{imbens1994identification}
Imbens, G.~W. and Angrist, J.~D. (1994).
\newblock {I}dentification and {E}stimation of {L}ocal {A}verage {T}reatment
  {E}ffects.
\newblock {\em {E}conometrica}, 62(2):467--475.

\bibitem[\protect\astroncite{Imbens and Rubin}{2015}]{imbens2015causal}
Imbens, G.~W. and Rubin, D.~B. (2015).
\newblock {\em Causal {I}nference in {S}tatistics, {S}ocial, and {B}iomedical
  {S}ciences}.
\newblock Cambridge University Press.

\bibitem[\protect\astroncite{Khan and Tamer}{2010}]{khan2010irregular}
Khan, S. and Tamer, E. (2010).
\newblock Irregular {I}dentification, {S}upport {C}onditions, and {I}nverse
  {W}eight {E}stimation.
\newblock {\em Econometrica}, 78(6):2021--2042.

\bibitem[\protect\astroncite{Kline and Walters}{2019}]{kline2019heckits}
Kline, P. and Walters, C.~R. (2019).
\newblock On {H}eckits, {LATE}, and {N}umerical {E}quivalence.
\newblock {\em Econometrica}, 87(2):677--696.

\bibitem[\protect\astroncite{Knaus et~al.}{2018}]{knaus2018better}
Knaus, M., Lechner, M., and Reimers, A. (2018).
\newblock For {B}etter or {W}orse? {T}he {E}ffects of {P}hysical {E}ducation on
  {C}hild {D}evelopment.
\newblock {\em IZA Discussion Paper}.

\bibitem[\protect\astroncite{Li et~al.}{2018}]{li2016balancing}
Li, F., Morgan, K.~L., and Zaslavsky, A.~M. (2018).
\newblock Balancing {C}ovariates via {P}ropensity {S}core {W}eighting.
\newblock {\em Journal of the American Statistical Association},
  113(521):390--400.

\bibitem[\protect\astroncite{Li et~al.}{2009}]{li2009efficient}
Li, Q., Racine, J.~S., and Wooldridge, J.~M. (2009).
\newblock {E}fficient {E}stimation of {A}verage {T}reatment {E}ffects with
  {M}ixed {C}ategorical and {C}ontinuous {D}ata.
\newblock {\em {J}ournal of {B}usiness \& {E}conomic {S}tatistics},
  27(2):206--223.

\bibitem[\protect\astroncite{Ouyang et~al.}{2009}]{ouyang2009nonparametric}
Ouyang, D., Li, Q., and Racine, J.~S. (2009).
\newblock {N}onparametric {E}stimation of {R}egression {F}unctions with
  {D}iscrete {R}egressors.
\newblock {\em {E}conometric {T}heory}, 25(1):1--42.

\bibitem[\protect\astroncite{Pohlmeier et~al.}{2016}]{pohlmeier2016simple}
Pohlmeier, W., Seiberlich, R., and Uysal, S.~D. (2016).
\newblock {A} {S}imple and {S}uccessful {S}hrinkage {M}ethod for {W}eighting
  {E}stimators of {T}reatment {E}ffects.
\newblock {\em Computational Statistics \& Data Analysis}, 100:512--525.

\bibitem[\protect\astroncite{Poterba et~al.}{1995}]{poterba1995401}
Poterba, J.~M., Venti, S.~F., and Wise, D.~A. (1995).
\newblock Do 401(k) {C}ontributions {C}rowd out other {P}ersonal {S}aving?
\newblock {\em Journal of Public Economics}, 58(1):1--32.

\bibitem[\protect\astroncite{Rubin}{1974}]{rubin1974estimating}
Rubin, D.~B. (1974).
\newblock {E}stimating {C}ausal {E}ffects of {T}reatments in {R}andomized and
  {N}onrandomized {S}tudies.
\newblock {\em {J}ournal of {E}ducational {P}sychology}, 66(5):688.

\bibitem[\protect\astroncite{Rubin}{2007}]{rubin2007design}
Rubin, D.~B. (2007).
\newblock The {D}esign versus the {A}nalysis of {O}bservational {S}tudies for
  {C}ausal {E}ffects: {P}arallels with the {D}esign of {R}andomized {T}rials.
\newblock {\em Statistics in {M}edicine}, 26(1):20--36.

\bibitem[\protect\astroncite{Vytlacil}{2002}]{vytlacil2002independence}
Vytlacil, E. (2002).
\newblock Independence, {M}onotonicity, and {L}atent {I}ndex {M}odels: {A}n
  {E}quivalence {R}esult.
\newblock {\em Econometrica}, 70(1):331--341.

\bibitem[\protect\astroncite{Zhao}{2019}]{zhao2019covariate}
Zhao, Q. (2019).
\newblock Covariate {B}alancing {P}ropensity {S}core by {T}ailored {L}oss
  {F}unctions.
\newblock {\em The Annals of Statistics}, 47(2):965--993.

\bibitem[\protect\astroncite{Zhao and Percival}{2017}]{zhao2017entropy}
Zhao, Q. and Percival, D. (2017).
\newblock Entropy {B}alancing is {D}oubly {R}obust.
\newblock {\em Journal of Causal Inference}, 5(1).

\bibitem[\protect\astroncite{Zubizarreta}{2015}]{zubizarreta2015stable}
Zubizarreta, J.~R. (2015).
\newblock Stable {W}eights that {B}alance {C}ovariates for {E}stimation with
  {I}ncomplete {O}utcome {D}ata.
\newblock {\em Journal of the American Statistical Association},
  110(511):910--922.

\end{thebibliography}
	\bibliographystyle{apa}	
\end{document}